%% file: main.tex
\documentclass[runningheads,envcountsame]{llncs}

\makeatletter
\RequirePackage[bookmarks,unicode,colorlinks=true]{hyperref}%
   \def\@citecolor{blue}%
   \def\@urlcolor{blue}%
   \def\@linkcolor{blue}%

\def\orcidID#1{\href{http://orcid.org/#1}{\smash{\protect\raisebox{-1.25pt}{\protect\includegraphics{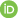}}}}}
\makeatother

\usepackage[T1]{fontenc}
\usepackage{xcolor,colortbl}
\usepackage{mathtools}
\usepackage{multirow}
\usepackage{stmaryrd}
\usepackage{cancel}
\usepackage{amsmath,amssymb}
\usepackage{thmtools, thm-restate}
\usepackage[shortlabels]{enumitem}
\usepackage{microtype}
\usepackage{wrapfig}
\usepackage{pbox}
\usepackage{marvosym}
\usepackage{listings}
\usepackage{enumitem}
\usepackage{ifthen}
\usepackage{stackengine}
\usepackage{xurl}
\hypersetup{
  breaklinks=true,
	pdftitle={Disproving (Positive) Almost-Sure Termination of Probabilistic Term Rewriting}, colorlinks=true, linkcolor=blue, citecolor=olive, filecolor=magenta, urlcolor=cyan
}

\usepackage[ruled,vlined,linesnumbered]{algorithm2e}

\SetCommentSty{CommentStyle}

\usepackage[most]{tcolorbox}
\tcbuselibrary{theorems}
\usepackage[capitalize,nameinlink]{cleveref}
\usepackage{IEEEtrantools}
\usepackage{placeins}
\usepackage{float}
\usepackage{tikz}
\usepackage[labelfont=bf]{caption}
\usepackage{subcaption}
\usepackage{booktabs}
\usepackage{nicefrac,xfrac}
\usepackage{centernot}
\usepackage{booktabs}
\usepackage{array,longtable}
\usetikzlibrary{shapes,calc,arrows,automata,decorations.pathmorphing,backgrounds,arrows.meta,shapes.geometric,shapes.multipart,shapes.misc,patterns,fit}
\pgfdeclarelayer{edgelayer}
\pgfdeclarelayer{nodelayer}
\pgfsetlayers{background,edgelayer,nodelayer,main}
\tikzstyle{none}=[inner sep=0mm]
\tikzstyle{moveBlock}=[fill=white, draw=black, shape=rectangle]
\tikzstyle{target}=[fill=white, draw=black, shape=circle]

\tikzstyle{dotHead}=[dotted, ->]
\tikzstyle{dotWithoutHead}=[dotted, -]
\tikzstyle{dashHead}=[dashed,->]
\tikzstyle{dashWithoutHead}=[dashed,-]
\tikzstyle{arrow}=[->]

\RequirePackage{makecell}

\newtcbtheorem{decisionProblem}{}{enhanced, fonttitle=\bfseries\upshape, fontupper=\slshape, arc=0mm, colback=white, colframe=black, rounded corners=all, opacityframe=0.8, coltitle=black, colbacktitle=white, boxsep=0.5mm, left=1mm, top=1mm, bottom=1mm, right=1mm}{decisionproblemcounter}

\newenvironment{myproof}{
	\noindent{\it Proof.}
}{\qed
	\medskip
}

\input{commands}

\definecolor{Gray}{gray}{0.85}
\definecolor{LightCyan}{rgb}{0.88,1,1}

\newcolumntype{a}{>{\columncolor{Gray}}c}
\newcolumntype{b}{>{\columncolor{white}}c}

 \newcommand{\makeproof}[2]{}
 \newcommand{\paper}[1]{}
 \newcommand{\report}[1]{#1}

 \newlength{\oldtextfloatsep}

\paper{\usepackage[ hyperref=true, backend=bibtex, firstinits=true, maxbibnames=99, sortcites, style=numeric-comp ]{biblatex}
\addbibresource{biblio.bib}}
\report{\usepackage[ hyperref=true, backend=bibtex, firstinits=true, maxbibnames=99, sortcites, style=numeric-comp ]{biblatex}
\addbibresource{biblio.bib}}

\title{Disproving (Positive) Almost-Sure Termination of Probabilistic Term Rewriting via Random Walks\thanks{funded by the DFG Research Training Group 2236 UnRAVeL}}
\titlerunning{Disproving Termination of PTRSs via Random Walks}
\author{J.-C.\
  Kassing\paper{$^{(\href{mailto:kassing@cs.rwth-aachen.de}{\mbox{\Letter}})}$}\orcidID{0009-0001-9972-2470}  
  \and H.\ Nagel\orcidID{0009-0001-9469-4618}  
  \and A.\ Schlecht\orcidID{0009-0004-3580-2764}  
  \and J.\ Giesl\paper{$^{(\href{mailto:giesl@informatik.rwth-aachen.de}{\mbox{\Letter}})}$}\orcidID{0000-0003-0283-8520}}
\institute{RWTH Aachen University, Aachen, Germany}
\authorrunning{J.-C.\ Kassing, H.\ Nagel, A.\ Schlecht, J.\ Giesl}

\begin{document}
\allowdisplaybreaks

\maketitle 
\input{abstract}
\input{introduction}
\input{pre}
\input{embedBasics}
\input{embedRW}
\input{embednonloops}
\input{conclusion}
\paper{\printbibliography}
\report{\printbibliography
  \appendix
  \input{appendix}
}
\end{document}

%% file: commands.tex
\makeatletter \newcommand*\bigcdot{\mathpalette\bigcdot@{.4}}
\newcommand*\bigcdot@[2]{\mathbin{\vcenter{\hbox{\scalebox{#2}{$\m@th#1\bullet$}}}}}
\makeatother

\renewcommand{\emptyset}{\varnothing}

\newcommand{\oldcomment}[1]{}
\renewcommand{\comment}[1]{\footnote{!!!#1}}
\renewcommand{\comment}[1]{}

\renewcommand{\epsilon}{\varepsilon}

\DeclareMathOperator{\dom}{dom}

\RequirePackage{tikz}             
\usetikzlibrary{shapes,calc,arrows,automata,decorations.pathmorphing,backgrounds,arrows.meta,shapes.geometric,shapes.multipart,shapes.misc}
\usetikzlibrary{arrows}
\usetikzlibrary{shapes.geometric}
\usetikzlibrary{arrows.meta}
\usetikzlibrary{calc}
\usetikzlibrary{intersections}
\usetikzlibrary{decorations.markings}
\usetikzlibrary{backgrounds}
\usetikzlibrary{matrix,arrows,decorations.pathmorphing}
\usetikzlibrary{positioning, 
                quotes}
\RequirePackage{xparse}
\RequirePackage{xspace}
\RequirePackage{tikz-cd}
\tikzset{commutative diagrams/.cd}
\RequirePackage{wrapfig}

\renewcommand{\emph}[1]{\index{#1}\textit{#1}}
\newcommand{\defemph}[1]{{\rm #1}}

\renewcommand{\emptyset}{\varnothing}

\newcommand{\IP}{\mathbb{P}}
\newcommand{\IZ}{\mathbb{Z}}
\newcommand{\IN}{\mathbb{N}}
\newcommand{\IR}{\mathbb{R}}
\newcommand{\expec}[1]{\ensuremath{\mathbb{E}\left(#1\right)}}
\newcommand{\IE}{\ensuremath{\mathbb{E}}}

\newcommand{\F}[1]{\mathfrak{#1}}

\newcommand{\pot}[1]{2^{#1}}

\makeatletter
\def\moverlay{\mathpalette\mov@rlay}
\def\mov@rlay#1#2{\leavevmode\vtop{%
   \baselineskip\z@skip \lineskiplimit-\maxdimen
   \ialign{\hfil$\m@th#1##$\hfil\cr#2\crcr}}}
\newcommand{\charfusion}[3][\mathord]{
    #1{\ifx#1\mathop\vphantom{#2}\fi
        \mathpalette\mov@rlay{#2\cr#3}
      }
    \ifx#1\mathop\expandafter\displaylimits\fi}
\makeatother

\newcommand{\aprove}{\textsf{AProVE}}

\newcommand{\ttttwo}{\textsf{T\kern-0.15em\raisebox{-0.55ex}T\kern-0.15emT\kern-0.15em\raisebox{-0.55ex}2}}

\newcommand{\tool}[1]{\textsf{#1}}

\newcommand{\TSet}[2]{\mathcal{T}\left(#1,#2\right)}

\newcommand{\VSet}{\mathcal{V}}

\newcommand{\R}{\mathcal{R}}

\newcommand{\FDist}{\operatorname{FDist}}
\newcommand{\Supp}{\operatorname{Supp}}

\newcommand{\rootterm}{\operatorname{root}}

\newcommand{\edl}{\operatorname{edl}}

\newcommand{\nonprob}{\normalfont{\text{np}}}

\newcommand{\nvd}{\normalfont{\text{nvd}}}
\newcommand{\PP}{\mathcal{P}}

\newcommand{\TT}{\mathcal{T}}

\newcommand{\tgt}{\mathsf{gt}}

\newcommand{\tz}{\mathsf{0}}

\renewcommand{\ts}{\mathsf{s}}

\newcommand{\tf}{\mathsf{f}}

\newcommand{\tg}{\mathsf{g}}

\newcommand{\ta}{\mathsf{a}}
\newcommand{\tb}{\mathsf{b}}
\newcommand{\tc}{\mathsf{c}}

\newcommand{\ttrue}{\mathsf{true}}
\newcommand{\tfalse}{\mathsf{false}}

\newcommand{\tgeo}{\mathsf{geo}}

\newcommand{\hookrule}{
  \mathrel{
    \xhookrightarrow{}
  }
}

\newcommand{\fun}[1]{\mathrm{#1}}
\newcommand{\pos}{\fun{Pos}}

\newcommand{\NF}{\mathtt{NF}} 
 
\newcommand{\ctroot}{\F{r}}
\newcommand{\pre}{\operatorname{Pre}}
\newcommand{\post}{\operatorname{Post}}
\newcommand{\ctleaf}{\operatorname{Leaf}}

\newcommand{\ctdepth}{\operatorname{d}}
\newcommand{\ctheight}{\operatorname{h}}

\newcommand{\AST}{\normalfont{\operatorname{\texttt{AST}}}}

\newcommand{\PAST}{\normalfont{\operatorname{\texttt{PAST}}}}

\newcommand{\PASTASTColor}{\normalfont{\operatorname{\texttt{\textcolor{purple}{(P)}AST}}}}

\makeatletter
\NewDocumentCommand{\dparrow}{+O{} +O{0.5cm}}{%
\begin{tikzpicture}[baseline=-0.5ex] {
\node[inner sep=0](@1) at (0,0) {};
\node[inner sep=0](@2) at (#2,0) {};
\draw [arrows={-Triangle[open]},shorten >= 1pt,shorten <= 1pt](@1)--(@2) node[pos=.5,above,inner sep=1pt] {\ensuremath{#1}};}
\end{tikzpicture}\xspace
}

\NewDocumentCommand{\myto}{+O{} +O{0.5cm}}{%
\begin{tikzpicture}[baseline=-0.5ex] {
\node[inner sep=0](@1) at (0,0) {};
\node[inner sep=0](@2) at (#2,0) {};
\draw [arrows={-to}](@1)--(@2) node[pos=.5,above,inner sep=1pt] {\ensuremath{#1}};}
\end{tikzpicture}\xspace
}

\NewDocumentCommand{\paraarrow}{+O{} +O{0.4cm}}{%
\begin{tikzpicture}[baseline=-0.5ex] {
\node[inner sep=0](@1) at (0,0) {};
\node[inner sep=0](@2) at (#2,0) {};
\node[inner sep=0](@3) at (0.07,0) {};
\draw [arrows={-to}](@1)--(@2) node[pos=.5,above,inner sep=1pt] {\ensuremath{#1}};
\draw [arrows={-to}](@1)--(@3);}
\end{tikzpicture}\xspace
}

\NewDocumentCommand{\paradparrow}{+O{} +O{0.4cm}}{%
\begin{tikzpicture}[baseline=-0.5ex] {
\node[inner sep=0](@1) at (0,0) {};
\node[inner sep=0](@2) at (#2,0) {};
\node[inner sep=0](@3) at (0.07,0) {};
\draw [arrows={-Triangle[open]}](@1)--(@2) node[pos=.5,above,inner sep=1pt] {\ensuremath{#1}};
\draw [arrows={-to}](@1)--(@3);}
\end{tikzpicture}\xspace
}

\newcommand{\oset}[2]{%
  {\mathop{#2}\limits^{\vbox to 1\ex@{\kern-\tw@\ex@
   \hbox{\scriptsize #1}\vss}}}}

\newcommand{\osetthree}[2]{%
  {\mathop{#2}\limits^{\vbox to 3\ex@{\kern-\tw@\ex@
   \hbox{\scriptsize #1}\vss}}}}

\newcommand{\osetfive}[2]{%
  {\mathop{#2}\limits^{\vbox to 5\ex@{\kern-\tw@\ex@
   \hbox{\scriptsize #1}\vss}}}}

\newcommand{\osetminus}[2]{%
  {\mathop{#2}\limits^{\vbox to -2\ex@{\kern-\tw@\ex@
   \hbox{\scriptsize #1}\vss}}}}
\makeatother

\newcommand{\subtermocAlt}{%
\mathrel{%
    \tikz[baseline=-0.5ex]{
      \node[inner sep=0pt, outer sep=0pt] (A) {$\blacktriangleleft$};
      \draw[white, line width=1.4pt]
        ($(A.north west)!0.515!(A.south west)$) --
        ($(A.north east)!0.515!(A.south east)$);
      \node[inner sep=0pt, outer sep=0pt] (A) {$\vartriangleleft$};
    }%
  }%
}

\newcommand{\geqMaybe}{%
  \mathrel{%
    \begin{tikzpicture}[scale=0.2, baseline=-0.9ex]
      \draw[line width=0.5pt] (0,0.5) -- (1,0) -- (0,-0.5);
      \draw[purple, line width=0.5pt] (0,-0.9) -- (1.1,-0.9);

      \draw[purple, -] (-0.1,-1.2) to[bend left] (-0.1,-0.6);
      \draw[purple, -] (1.2,-1.2) to[bend right] (1.2,-0.6);
    \end{tikzpicture}
  }
}

\newcommand{\NOO}{\operatorname{NO}}

\newcommand{\maxNOO}{\operatorname{maxNO}}
\newcommand{\maxParaNOO}{\operatorname{maxOO}}

\newcommand{\maxParaNOM}{\operatorname{maxOPO}}

\crefname{definition}{Def.}{Def.}
\crefname{example}{Ex.}{Ex.}
\crefname{counterexample}{Counterex.}{Counterex.}
\crefname{appendix}{App.}{App.}
\crefname{ex}{Ex.}{Ex.}
\crefname{theorem}{Thm.}{Thm.}
\crefname{lemma}{Lemma}{Lemmas}
\crefname{remark}{Remark}{Remarks}
\crefname{section}{Sect.}{Sect.}
\crefname{subsection}{Sect.}{Sect.}
\crefname{subsubsection}{Sect.}{Sect.}
\crefname{line}{Line}{Lines}
\crefname{corollary}{Cor.}{Cor.}
\crefname{figure}{Fig.}{Fig.}
\crefname{enumi}{}{}
\crefname{algorithm}{Alg.}{Alg.}

%% file: abstract.tex
\begin{abstract}
  While numerous techniques have been developed to automatically prove termination of
   probabilistic programs, there are only few
  automated methods to \emph{disprove} their
  termination. In this paper, we present the first techniques to
  automatically disprove (positive) almost-sure termination of probabilistic term rewriting.
  Disproving termination of non-probabilistic systems requires finding 
  a finite representation of an infinite computation, e.g., a loop of the rewrite system.
  We extend such qualitative techniques to probabilistic term rewriting, 
  where a quantitative analysis is required.
  In addition to the existence of a loop, we have to count the number of such loops in
  order to
  embed suitable random walks into a computation, thereby disproving termination. 
  To evaluate their power, we implemented all our techniques in the tool \aprove{}.
\end{abstract}

%% file: introduction.tex
\section{Introduction}\label{sec-introduction}

While termination is undecidable in general, automatic techniques to prove termination are
crucial in many applications. However, proving \emph{non-termination} is equally
important, e.g., to 
detect bugs.
While (non-)termination of ordinary programs
has been studied for decades, the work on automated termination analysis of
\emph{probabilistic programs} is quite recent.
Based on \emph{probabilistic choices}, 
such programs may proceed in several different ways.
Probabilistic programming is used to deal with
uncertainty in data and
has many applications, see \cite{Gordon14, barthe2020FoundationsProbabilisticProgramming}.

In the probabilistic setting,  one usually considers notions 
like (positive) almost-sure termination.
A program is \emph{almost-surely terminating}
($\AST$) if every computation terminates with probability $1$. 
A strictly stronger notion is \emph{positive} $\AST$ ($\PAST$) \cite{bournez2005proving,DBLP:conf/mfcs/Saheb-Djahromi78},
where every computation must have a finite expected number of steps.
It is well known that $\PAST$ implies $\AST$ but not vice versa.

Term rewriting \cite{baader_nipkow_1999} is a fundamental concept to transform and
evaluate expressions, and techniques to analyze termination of term rewrite systems (TRSs)
are used for termination analysis of programs in many languages.
Term rewriting was adapted to the probabilistic setting in
\cite{BournezRTA02,bournez2005proving,avanzini2020probabilistic}.
While techniques to automatically prove $\AST$ and $\PAST$  of
probabilistic TRSs (PTRSs) were
developed in \cite{avanzini2020probabilistic,kassinggiesl2023iAST,FoSSaCS2024,kassing2025DependencyPairsExpecteda,kassing2026AnnotatedDependencyPaira}, up
to now  there were no approaches for non-termination of PTRSs.
In this paper, we develop the first techniques to automatically disprove  $\AST$ and $\PAST$ of PTRSs.

\smallskip

\textbf{Contribution.} 
We introduce a general approach to disprove termination of PTRSs 
by embedding \paper{\pagebreak[3]}
non-terminating random walks into rewrite sequences of PTRSs (\Cref{theorem:lower_bounds_by_embeddings}).
First, we embed random walks via \emph{occurrences} of \emph{loops} 
(\Cref{theorem:embedding}, \ref{theorem:embedding_two}, and \ref{theorem:embedding_three}), 
which were previously used to disprove termination of TRSs, see, e.g.,\linebreak[3] \cite{Frocos05}.
Second, we embed random walks via looping \emph{pattern terms} (\Cref{theorem:embedding-pattern}),
which were previously used to prove non-looping non-termination 
of TRSs  \cite{emmes2012ProvingNonloopingNontermination}.
While the original techniques were \emph{qualitative} 
(\emph{find the existence of a loop}), 
the probabilistic setting requires \emph{quantitative} analyses 
(\emph{find the number of possible loops}).
We develop a dynamic programming algorithm (\Cref{alg:maxNOO}) 
to solve the counting problems that arise in this setting,
e.g., to determine how many instantiations of a term $t$ occur in a term $s$ 
at orthogonal positions.
We implemented our techniques in the tool \textsf{AProVE} 
and demonstrate their applicability by an experimental evaluation.

\smallskip

\textbf{Related Work.}
There exist several techniques to prove non-termination of non-probabilistic term rewriting, see, e.g.,
\cite{Frocos05,emmes2012ProvingNonloopingNontermination,payet2008LoopDetectionTerm,payet2024NonterminationTermRewriting,DBLP:conf/rta/EndrullisZ15}.
Tools analyzing (non-)termination of TRSs participate annually 
in the \emph{Termination Competition} (\textit{TermComp}) \cite{termcomp}.
A related problem to proving non-termination is
analyzing reachability (or (in)feasibility) for TRSs, see, e.g.,
\cite{sternagelReachabilityAnalysisTermination2019,
  lucas2018UseLogicalModels, 
  gutierrez2020AutomaticallyProvingDisproving, yamadaTermOrderingsNonreachability2022a}.

For imperative programs, there are also numerous
techniques for proving non-ter\-mination, e.g.,
\cite{gupta2008ProvingNontermination, brockschmidt2012AutomatedDetectionNontermination,
chen2014ProvingNonterminationSafety, larraz2014ProvingNonterminationUsing, 
leike2018GeometricNonterminationArguments, frohn2022CalculusModularLoop, 
frohn2023ProvingNonTerminationAcceleration}, and tools for termination analysis of
imperative programs compete annually in \textit{TermComp} and
in the software verification\linebreak competition (\textit{SV-COMP})
\cite{beyer2025ImprovementsSoftwareVerification}.
There also exist \emph{decision} procedures for
termination on certain subclasses of programs, e.g.,
\cite{tiwari2004TerminationLinearPrograms, braverman2006TerminationIntegerLinear, xu2013SymbolicTerminationAnalysis,  hosseini2019TerminationLinearLoops, 
  FMSD2025, TargetingCompletenessJAR}.
To automatically disprove $\PASTASTColor$,
we are only aware of approaches for probabilistic \emph{imperative} programs:
The technique of 
\cite{chatterjee2017StochasticInvariantsProbabilistic}
works via repulsing supermartingales,
and the technique of \cite{takisaka2021RankingRepulsingSupermartingales} uses
fixpoints and Park induction.
Based on \cite{chatterjee2017StochasticInvariantsProbabilistic}, 
martingale-based\linebreak[3] proof rules which can also disprove $\AST$ and $\PAST$ were implemented in the
tool \textsf{Amber} \cite{amber}.
The hardness of analyzing $\AST$ and $\PAST$ was investigated in \cite{kaminski2019hardness,majumdarPositiveAlmostSureTermination2024}. 

The most common non-termination technique for TRSs is the detection of loops.
Thus, to disprove $\AST$ or $\PAST$ of probabilistic term rewriting,
in this work we lift the idea of disproving termination via loops
to the probabilistic setting, and consider embeddings of random walks based on loops.

In \cite{beutner2021probabilistic},
random walks are also used to
analyze $\AST$ (of higher-order probabilistic functional programs). However,
here one
over-approximates all runs by 
random walks in order to prove $\AST$ by counting recursive calls.
Instead, our goal is to under-approximate a possible run in order to disprove $\AST$.
This requires additional conditions to ensure that
the counted calls are indeed executable.

\smallskip

\textbf{Structure.} We present preliminaries on probability theory, 
random walks, and (probabilistic) term rewriting in \Cref{Preliminaries}.
Then, we introduce our novel approach to disprove termination of PTRSs 
via an embedding of random walks in \Cref{Embedding Random Walks}.
More precisely,
in \Cref{Occ Random Walks}
we show how to embed random walks based on loops and 
develop a dynamic programming algorithm
to solve the arising counting problems for occurrences of terms. 
In \Cref{Pattern Random Walks}, 
we show how to embed random walks based on looping pattern terms. 
We evaluate our implementation and  conclude in \Cref{Conclusion}.\report{
The proofs of our results can be found in App.\ \ref{Appendix}.}\paper{ All proofs can be
  found in \cite{report}.}

%% file: pre.tex
\section{Preliminaries}\label{Preliminaries}

In \Cref{Probability Theory}, \paper{\vspace*{-.3cm}\pagebreak}
we start with basic concepts of probability theory 
and define the notion of 
 \emph{random walk programs (without direct termination)} as in
\cite{giesl2019ComputingExpectedRuntimes}, which we will use as lower bounds to disprove
(positive) almost-sure termination of probabilistic term rewrite systems in \Cref{Occ Random Walks} and \ref{Pattern Random Walks}.
For an introduction to general random walks, see, e.g., \cite{spitzer2001, lawler2010RandomWalkModern, grimmett2020probability}.
We recapitulate ordinary term rewriting in \Cref{sec:TRS}, and probabilistic term rewriting in \Cref{sec:PTRS}.

\subsection{Probability Theory and Random Walks}\label{Probability Theory}

In probability theory, one considers
\emph{outcomes} $\alpha \in \Omega$ of a random process and
 sets of \emph{events} $\F{A} \subseteq \pot{\Omega}$ 
(where an event is a subset of outcomes),
and one assigns probabilities to each event.
Such an \emph{event space} $\F{A}$ has to contain $\Omega$ 
and it must be closed under complement and countable unions.
The pair $(\Omega, \F{A})$ is called a \emph{measurable space}.
A \emph{probability space} $(\Omega, \F{A}, \IP)$ extends a measurable space $(\Omega, \F{A})$
by a \emph{probability measure} $\IP$ which maps every event from $\F{A}$ 
to a probability between $0$ and $1$ such that $\IP(\Omega) = 1$, $\IP(\emptyset) = 0$,
and $\IP(\biguplus_{i \geq 0} A_i) = \sum_{i \geq 0} \IP(A_i)$ for 
pairwise disjoint
events $A_i \in \F{A}$.
As in \cite{giesl2019ComputingExpectedRuntimes},
we use the measurable space $(\IZ^{\omega}, \F{A}_{\text{cyl}})$ on infinite words 
defined by the typical cylinder construction used in MDP theory.
We call the possible outcomes $\alpha \in \IZ^{\omega}$ \emph{runs}
and a finite word $\alpha \in \IZ^{*}$ a \emph{prefix run}.
Let $\alpha(i)$ be the number at position $i$ in $\alpha$.

A \emph{random walk (program)}
$\mu$ is a function $\mu: \IZ \to \IR_{\geq 0}$
whose \emph{support} $\Supp(\mu) = \{x \in \IZ \mid \mu(x) > 0\}$ is finite and satisfies
$\sum_{x \in \Supp(\mu)} \mu(x) = 1$.
The function $\mu$ represents the transition relation of a random walk on $\IZ$.
If the current value of the random walk is $y > 0$, then for any $x \in \IZ$,
$\mu(x)$ is the probability that the random walk transitions from $y$ to $y+x$.
We stop when we reach a number\linebreak[3] $\leq 0$ and do not perform any transition steps
anymore.\footnote{Ordinary random walks  \cite{spitzer2001} do not stop and are
identically distributed everywhere.}
For a prefix run $\alpha \in \IZ^{*}$, let 
$\langle \alpha \rangle \subseteq \IZ^{\omega}$ denote the set of all infinite words 
with prefix $\alpha$ (called a \emph{cylinder set}).
The
probability measure $\IP^{\mu}_{x_0}$ (given some start value $x_0 \in \IZ$) 
of a random walk is defined on the
event space $\F{A}_{\text{cyl}} = \{ \langle \alpha
\rangle \mid \alpha \in  \IZ^{*} \}$, i.e., on the
cylinder sets $\langle \alpha \rangle$ 
of all prefix runs $\alpha$.
Given a finite word $\alpha = \alpha(0) \ldots \alpha(k)$,
we define $\IP^{\mu}_{x_0}(\langle \alpha \rangle) = 0$ if $\alpha(0) \neq x_0$,
and $\IP^{\mu}_{x_0}(\langle \alpha \rangle) = \prod_{i = 0}^{k-1} \mu(\alpha(i+1) -
\alpha(i))$ otherwise.

Depending on $\mu$ and its \emph{expected value} $\expec{\mu} = \sum_{x \in \Supp(\mu)} x
\cdot \mu(x)$, we characterize four different types of random walks:
\begin{enumerate}
    \item If $\mu(0) = 1$, then $\mu$ is a \emph{loop walk}.
    \item If $\mu(0) < 1$ and $\expec{\mu} = 0$, then $\mu$ is a \emph{symmetric random walk}.
    \item If $\mu(0) < 1$ and $\expec{\mu} > 0$, then $\mu$ is a \emph{positively biased random walk}.
    \item If $\mu(0) < 1$ and $\expec{\mu} < 0$, then $\mu$ is a \emph{negatively biased random walk}.
\end{enumerate}
\Cref{fig:all} shows examples for all four different types, where we start at
$x_0 = 1$.

\begin{figure}[t]
  \centering
    \scriptsize
    \begin{subfigure}{0.47\textwidth}
        \centering
        \begin{tikzpicture}[scale=1.0]
            \draw (4.5,0.5) node{$\begin{aligned}
                \phantom{\mu_1(-1) =\;}& \phantom{\nicefrac{1}{3}} \\
                \mu_1(0) =\;& 1 \\
                \phantom{\mu_1(1) =\;}& \phantom{\nicefrac{1}{3}} \end{aligned}$};
                
            \draw (0,0) --++(90:1cm);
            \fill[pattern=north west lines] (0,0) rectangle ++(-0.5,1); 

            \draw[->, thick] (-0.5,0) -- (3,0);
            \foreach \x/\den in {0/50, 1/50, 2/50} {
                \node[circle, fill=gray!\den, inner sep=0pt, minimum size=0.35cm] (\x-a) at (\x,0) {\x};
            }

            \draw[-Stealth, dashed, thick, yshift=-7pt] (1-a) to[loop above, distance=9mm, in=70, out=110] node[pos=0.5, above, xshift=5pt] {$1$} (1-a);
        \end{tikzpicture}
        \caption{Loop Walk $\mu_1$}
        \label{fig:sub1}
    \end{subfigure}
    \hfill
    \begin{subfigure}{0.47\textwidth}
        \centering
        \begin{tikzpicture}[scale=1.0]
            \draw (4.5,0.5) node{$\begin{aligned}
            \mu_2(-1) =\;& \nicefrac{1}{3} \\
            \mu_2(0) =\;& \nicefrac{1}{3} \\
            \mu_2(1) =\;& \nicefrac{1}{3} \end{aligned}$};

            \draw (0,0) --++(90:1cm);
            \fill[pattern=north west lines] (0,0) rectangle ++(-0.5,1); 

            \draw[->, thick] (-0.5,0) -- (3,0);
            \foreach \x/\den in {
                0/50, 1/50, 2/50} {
                \node[circle, fill=gray!\den, inner sep=0pt, minimum size=0.35cm] (\x-b) at (\x,0) {\x};
            }

            \draw[-Stealth, dashed, thick, yshift=3pt, shorten >= 2pt, shorten <= 2.5pt] (1,0) to[bend left] node[above] {$\nicefrac{1}{3}$} (2,0);
            \draw[-Stealth, dashed, thick, yshift=-7pt] (1-b) to[loop above, distance=9mm, in=70, out=110] node[pos=0.5, above, xshift=5pt] {$\nicefrac{1}{3}$} (1-b);
            \draw[-Stealth, dashed, thick, yshift=3pt, shorten >= 2pt, shorten <= 2.5pt] (1,0) to[bend right] node[above] {$\nicefrac{1}{3}$} (0,0);
        \end{tikzpicture}
        \caption{Symmetric Random Walk $\mu_2$}
        \label{fig:sub2}
    \end{subfigure}
    \begin{subfigure}{0.47\textwidth}
        \centering
        \begin{tikzpicture}[scale=1.0]
            \draw (4.5,0.5) node{$\begin{aligned}
            \mu_3(-1) =\;& \nicefrac{1}{3} \\
            \mu_3(1) =\;& \nicefrac{2}{3} \end{aligned}$};

            \draw (0,0) --++(90:1cm);
            \fill[pattern=north west lines] (0,0) rectangle ++(-0.5,1); 

            \draw[->, thick] (-0.5,0) -- (3,0);
            \foreach \x/\den in {
                0/50, 1/50, 2/50} {
                \node[circle, fill=gray!\den, inner sep=0pt, minimum size=0.35cm] at (\x,0) {\x};
            }
        
            \draw[-Stealth, dashed, thick, yshift=3pt, shorten >= 2pt, shorten <= 2.5pt] (1,0) to[bend left] node[above] {$\nicefrac{2}{3}$} (2,0);
            \draw[-Stealth, dashed, thick, yshift=3pt, shorten >= 2pt, shorten <= 2.5pt] (1,0) to[bend right] node[above] {$\nicefrac{1}{3}$} (0,0);
        \end{tikzpicture}
        \caption{Positively Biased Random Walk $\mu_3$}
        \label{fig:sub3}
    \end{subfigure}
    \hfill
    \begin{subfigure}{0.47\textwidth}
        \centering
        \begin{tikzpicture}[scale=1.0]
            \draw (4.5,0.5) node{$\begin{aligned}
            \mu_4(-1) =\;& \nicefrac{2}{3} \\
            \mu_4(1) =\;& \nicefrac{1}{3} \end{aligned}$};

            \draw (0,0) --++(90:1cm);
            \fill[pattern=north west lines] (0,0) rectangle ++(-0.5,1); 

            \draw[->, thick] (-0.5,0) -- (3,0);
            \foreach \x/\den in {
                0/50, 1/50, 2/50} {
                \node[circle, fill=gray!\den, inner sep=0pt, minimum size=0.35cm] at (\x,0) {\x};
            }
        
            \draw[-Stealth, dashed, thick, yshift=3pt, shorten >= 2pt, shorten <= 2.5pt] (1,0) to[bend left] node[above] {$\nicefrac{1}{3}$} (2,0);
            \draw[-Stealth, dashed, thick, yshift=3pt, shorten >= 2pt, shorten <= 2.5pt] (1,0) to[bend right] node[above] {$\nicefrac{2}{3}$} (0,0);
        \end{tikzpicture}
        \caption{Negatively Biased Random Walk $\mu_4$}
        \label{fig:sub4}
    \end{subfigure}

    \caption{Four Different Random Walks $\mu_1$, $\mu_2$, $\mu_3$, and $\mu_4$}
    \label{fig:all}
    \vspace*{-10px}
\end{figure}

We are interested in the probability of termination 
and the expected number of steps it takes to terminate. 
The random walk $\mu$ \emph{certainly terminates} if
there is no infinite word $\alpha$ with
$\alpha(i) >0$ and
$\mu(\alpha(i+1) - \alpha(i)) > 0$
for all $i \in \IN$.  
This only holds if $\mu(x) = 0$ for all $x \geq 0$. 
Since this is a rather restrictive requirement,
we are more interested in the \emph{probability of termination}.
A run $\alpha$ \emph{terminates} if there exists an $n \in \IN$ such that
$\alpha(n) \leq 0$,
and the termination length $|\alpha|$ is defined as the smallest such $n$.  \pagebreak[3]
Let $\langle \mathtt{SN} \rangle = \biguplus_{n \in \IN} \biguplus_{\alpha \in \IZ^{\omega}, |\alpha| = n} \langle \alpha \rangle$
be the event of all terminating runs. 
(As usual, ``\texttt{SN}'' stands for ``\underline{s}trong \underline{n}ormalization'',
which is equivalent to ``termination''.)
Then the \emph{probability of termination} of a random walk $\mu$ starting in $x_0$ is 
$\IP^{\mu}_{x_0}(\langle \mathtt{SN} \rangle) 
= \sum_{n \in \IN} \sum_{\alpha \in \IZ^{\omega}, |\alpha| = n}
\IP^{\mu}_{x_0}(\langle \alpha \rangle)$.
  
A random walk $\mu$ is \emph{almost-surely terminating} ($\AST$) 
if $\IP^{\mu}_{x_0}(\langle \mathtt{SN} \rangle) = 1$ for all $x_0 \in \IZ$.
In addition, one is also interested in the expected number of steps it takes to terminate.
We define the \emph{expected derivation length} $\edl(\mu, x_0) \in \IR_{\geq 0} \cup \{ \infty \}$ 
of a random walk $\mu$ starting in $x_0$ as $\edl(\mu, x_0) = \infty$
if $\mu$ is not $\AST$, and otherwise, as
$\edl(\mu, x_0) = \sum_{n \in \IN} \sum_{\alpha \in \IZ^{\omega}, |\alpha| = n} n
\cdot \IP^{\mu}_{x_0}(\langle \alpha \rangle)$.
A random walk $\mu$ is \emph{positively almost-surely terminating} ($\PAST$) 
if we have $\edl(\mu, x_0) < \infty$ for all $x_0 \in \IZ$.

With this characterization, as in \cite[Thm.\ 18]{giesl2019ComputingExpectedRuntimes},
classical results on random walks \cite{spitzer2001} yield the following
classification w.r.t.\ $\AST$ and $\PAST$.

\begin{theorem}[$\AST$ and $\PAST$ of Random Walks]\label{theorem:termination}
    Let $\mu$ be a random walk.
    \begin{enumerate}
        \item If $\mu$ is a loop walk, then $\mu$ is neither $\AST$ nor $\PAST$.
        \item If $\mu$ is a symmetric random walk, then $\mu$ is $\AST$ but not $\PAST$.
        \item If $\mu$ is a positively biased random walk, then $\mu$ is neither $\AST$ nor $\PAST$.
        \item If $\mu$ is a negatively biased random walk, then $\mu$ is $\AST$ and $\PAST$.
    \end{enumerate}
\end{theorem}

\begin{example}\label{ex:terminationprob}
    Based on \Cref{theorem:termination}, 
    $\mu_1$ and $\mu_3$ are neither $\AST$ nor $\PAST$.
    The random walk $\mu_2$ is $\AST$ but not $\PAST$, 
    and $\mu_4$ is both $\AST$ and $\PAST$.
\end{example}

\subsection{Rewriting}\label{sec:TRS}

Next, we recapitulate
abstract rewriting and term rewriting \cite{baader_nipkow_1999}.
An \emph{abstract reduction system} (ARS) on a set $A$ is a binary relation $\to \; \subseteq A
\times A$.
Instead of $(a,b) \in \; \to$ one also writes $a \to b$.
For any  $n \in \IN$, we define $\to^{n}$ as 
$\to^{0} \;=\; \{(a,a) \mid a \in A\}$ and $\to^{n+1} \;=\; \to^{n} \circ \to$, 
where ``$\circ$'' denotes composition of relations. 
Moreover, $\to^* = \bigcup_{n \in \IN}\to^{n}$
and $\to^+ = \bigcup_{n \in \IN_{>0}}\to^{n}$.
$\NF_{\to}$ denotes the set of all objects  in \emph{normal form}
w.r.t.\ $\to$, i.e., $a \in \NF_{\to}$ if there is no $b \in A$ with  $a \to b$. 

The set $\TSet{\Sigma}{\VSet}$ of all \emph{terms}
over a finite set of \emph{function symbols} $\Sigma = \biguplus_{k \in \IN}\Sigma_k$
and an infinite set of \emph{variables} $\VSet$
is the smallest set with $\VSet \subseteq \TSet{\Sigma}{\VSet}$, 
and if $f \in \Sigma_k$
and $t_1, \dots, t_k \in \TSet{\Sigma}{\VSet}$,
then $f(t_1,\dots,t_k) \in \TSet{\Sigma}{\VSet}$.
If $\Sigma$ and $\VSet$ are clear, we just write $\TT$ instead of
$\TSet{\Sigma}{\VSet}$.
For example, $\tgt(\ts(x),\tz)$ is a
term over a signature $\Sigma$ where
$\tgt \in \Sigma_2$, $\ts \in \Sigma_1$, and $\tz \in \Sigma_0$.
A \emph{substitution} is a 
function $\sigma:\VSet \to \TSet{\Sigma}{\VSet}$ \pagebreak[3] where $\dom(\sigma) = \{x \in \VSet \mid \sigma(x) \neq x\}$ is finite, 
and we often write $x\sigma$ instead of $\sigma(x)$.
If $\dom(\sigma) = \{ x_1,\ldots,x_n\}$ and $\sigma(x_i) = s_i$ 
for all $1 \leq i \leq n$,  we
also write $\sigma = [x_1/s_1, \ldots, x_n/s_n]$.
Substitutions homomorphically extend 
to terms: if $t=f(t_1,\dots,t_k)\in \TT$ then $t \sigma
= f(t_1\sigma,\dots,t_k \sigma)$.
Thus, for a substitution $\sigma$ with $x \sigma = \ts(\tz)$ we obtain $\ts(x) \sigma
= \ts(\ts(\tz))$. 
For any term $t \in \TT$, the set of \emph{positions} $\pos(t)$ 
is the smallest subset of $\IN^*$ with $\varepsilon \in \pos(t)$, 
and if $t=f(t_1,\dots,t_k)$ then for all $1 \leq j \leq k$  and all $\pi \in \pos(t_j)$ we
have $j.\pi \in \pos(t)$.  
$\pos_{X}(t) \subseteq \pos(t)$  denotes all positions 
of symbols or variables from $X \subseteq \Sigma \cup \VSet$. 
A position $\pi_1$ is \emph{above}
$\pi_2$ if $\pi_1$ is a (not necessarily proper)
prefix of $\pi_2$.
If $\pi_1$ is not above $\pi_2$ and $\pi_2$ is not above $\pi_1$,
then they are \emph{orthogonal} (denoted $\pi_1 \bot \pi_2$).
If $\pi \in \pos(t)$ then $t|_{\pi}$ denotes the subterm  at position $\pi$
and $t[r]_{\pi}$ denotes the term that results from replacing the subterm $t|_{\pi}$ at position $\pi$ with the term $r \in \TT$.
We write $t \trianglelefteq s$ if $t$ is a subterm of $s$ and $t \vartriangleleft s$ 
if $t$ is a \emph{proper} subterm of $s$ (i.e., if $t \trianglelefteq s$ and $t \neq s$).
For example, we have $\pos(\tgt(\ts(x),\tz)) = \{\varepsilon, 1, 1.1, 2\}$, $\tgt(\ts(x),\tz)|_{2} = \tz$, 
$\tgt(\ts(x),\tz)[\ts(y)]_{2} = \tgt(\ts(x),\ts(y))$, and $\ts(y) \vartriangleleft \tgt(\ts(x),\ts(y))$.
A \emph{context} $C$ is a term from
$\TSet{\Sigma \uplus \{ \Box \}}{\VSet}$ 
which contains exactly one occurrence
of the constant $\Box$  (called ``hole'').
If $C|_\pi = \Box$, then $C[s]$ is a shorthand for $C[s]_\pi$.

A \emph{term rewrite rule} $\ell \to r$ is a pair of terms $(\ell, r) \in \TT \times \TT$ 
with $\VSet(r) \subseteq \VSet(\ell)$ and $\ell \notin \VSet$,
where $\VSet(t)$ is the set of all variables occurring in $t \in \TT$.
A \emph{term rewrite system} (TRS) $\R$ is a finite set of term rewrite rules,
and it induces an ARS $(\TT, \to_{\R})$ where $s \to_{\R} t$ 
holds if there is a  $\pi \in \pos(s)$, 
a rule $\ell \to r \in \R$, and a substitution $\sigma$ 
with $s|_{\pi}=\ell\sigma$ and $t = s[r\sigma]_{\pi}$.
We sometimes simply refer to $\R$ instead of $\to_{\R}$.
For example, the following TRS $\R_{\tgt}$
computes the ``\textbf{g}reater \textbf{t}han'' function on natural numbers
(represented by $\tz$ and the successor function $\ts$).
\[ \begin{array}{@{}r@{\;}c@{\;}l@{\qquad}r@{\;}c@{\;}l@{\qquad}r@{\;}c@{\;}l}
  \tgt(\ts(x),\ts(y)) &\to& \tgt(x,y) & \tgt(\ts(x),\tz) &\to& \ttrue & \tgt(\tz,y) &\to& \tfalse
\end{array}\]
Here, we have
$\tgt(\ts(\tz),\ts(\tz)) \to_{\R_{\tgt}} \tgt(\tz,\tz) \to_{\R_{\tgt}} \tfalse$, 
where $\tfalse \in \NF_{\R_{\tgt}}$.

\subsection{Probabilistic Rewriting}\label{sec:PTRS}

A \emph{probabilistic} ARS has finite multi-distributions\footnote{We use
multi-distributions instead of ordinary distributions. Here, the same object can occur multiple times and, due to non-determinism, be evaluated differently.} on the right-hand sides of its rewrite rules.
A finite \emph{multi-distribution} $\mu$ on a set $A \neq \emptyset$ is a 
finite multiset of pairs $(p:a)$, with a probability $0 < p \leq 1$ 
and $a \in A$, such that $\sum _{(p:a) \in \mu} p = 1$.
$\FDist(A)$ denotes the set of all finite multi-distributions on $A$.
For $\mu\in\FDist(A)$, its \emph{support} is the multiset $\Supp(\mu)\!=\!\{a \mid (p\!:\!a)\!\in\!\mu$ for some $p\}$.
A \emph{probabilistic abstract reduction system} (PARS) is a pair $(A, \to)$ with
$\to \; \subseteq A \times \FDist(A)$.

A \emph{probabilistic term rewrite rule} $\ell \to \mu$ is a pair $(\ell, \mu) \in \TT \times \FDist(\TT)$ 
such that $\ell \not\in \VSet$ and $\VSet(r) \subseteq \VSet(\ell)$ for all $r \in \Supp(\mu)$,
and a \emph{probabilistic TRS} (PTRS) is a finite set of probabilistic term rewrite
rules. 
Similar to TRSs, a PTRS $\PP$ induces a PARS $(\TT, \to_{\PP})$ with
$s \to_{\PP} \{p_1:t_1, \ldots, p_k:t_k\}$  
if there is a position $\pi \in \pos(s)$, a rule $\ell \to \{p_1:r_1, \ldots, p_k:r_k\} \in \PP$, 
and a substitution $\sigma$ such that $s|_{\pi}=\ell\sigma$ 
and $t_j = s[r_j\sigma]_{\pi}$ for all $1 \leq j \leq k$.
Again, we sometimes refer to $\PP$ instead of $\to_{\PP}$.
Consider the PTRS $\PP_{\tgeo}$ with the only rule $\tgeo(x) \to
\{\nicefrac{1}{2}:\tgeo(\ts(x)), \; \nicefrac{1}{2}:x \}$.
When starting with the term $\tgeo(\tz)$, repeated rewriting
yields $\ts^k(\tz)$ (representing $k \in \IN$) with a probability of $(\nicefrac{1}{2})^{k+1}$, 
i.e., \pagebreak[3] a geometric distribution.

To track rewrite sequences of a PARS $(A, \to)$ with their probabilities,
we consider (possibly infinite) \emph{rewrite sequence trees (RSTs)}~\cite{kassing2026AnnotatedDependencyPaira}.
The nodes $v$ of a $\to$-RST are labeled by pairs $(p_v:a_v)$ of a
probability $p_v \in (0,1]$ and an object $a_v \in A$, where
the probability at the root is $1$. 
For each node $v$ with successors $w_1, \ldots, w_k$, 
the edge relation represents a rewrite step,
i.e., $a_v \to \{\tfrac{p_{w_1}}{p_v}:a_{w_1}, \ldots, \tfrac{p_{w_k}}{p_v}:a_{w_k}\}$.
For a $\to$-RST $\F{T}$,
$V(\F{T})$ denotes its set of nodes, $\rootterm(\F{T})$ is the object at its root, 
and $\ctleaf(\F{T})$ denotes its set of leaves.
The \emph{depth} $\ctdepth(v)$ of a node $v \in V(\F{T})$ is 
the number of steps it takes to reach $v$ from the root. 
The \emph{height} of a tree $\F{T}$ is $\ctheight(\F{T}) = \sup_{v \in V(\F{T})} \ctdepth(v) \in \IN \cup \{\omega\}$.
An example for a ${\PP_{\tgeo}}$-RST is shown in \Cref{fig:RST_example}.

\begin{figure}[t]
  \centering
    \scriptsize
    \begin{tikzpicture}
        \tikzstyle{adam}=[thick,draw=black!100,fill=white!100,minimum size=4mm,
          shape=rectangle split, rectangle split parts=2,rectangle split horizontal,font={\scriptsize}]
        \tikzstyle{empty}=[rectangle,thick,minimum size=4mm]
        
        \node[adam] at (0, 0)  (a) {$1$\nodepart{two}$\tgeo(\tz)$};

        \node[adam] at (3, 0)  (b) {$\nicefrac{1}{2}$\nodepart{two}$\tgeo(\ts(\tz))$};
        \node[adam,label=right:{\footnotesize $\NF_{\PP_{\tgeo}}$}] at (3, -0.7)  (c) {$\nicefrac{1}{2}$\nodepart{two}$\tz$};

        \node[adam] at (6, 0)  (d) {$\nicefrac{1}{4}$\nodepart{two}$\tgeo(\ts(\ts(\tz)))$};
        \node[adam,label=right:{\footnotesize $\NF_{\PP_{\tgeo}}$}] at (6, -0.7)  (e) {$\nicefrac{1}{4}$\nodepart{two}$\ts(\tz)$};

        \node[empty] at (9, 0)  (f) {$\ldots$};
        \node[adam,label=right:{\footnotesize $\NF_{\PP_{\tgeo}}$},label=above:{\footnotesize \textcolor{purple}{$v$}}] at (9, -0.7)  (g) {$\nicefrac{1}{8}$\nodepart{two}$\ts(\ts(\tz))$};
        
        \draw (a) edge[->] (b);
        \draw (a) edge[->] (c);
        \draw (b) edge[->] (d);
        \draw (b) edge[->] (e);
        \draw (d) edge[->] (f);
        \draw (d) edge[->] (g);
    \end{tikzpicture}\vspace*{-.1cm}
    \caption{A ${\PP_{\tgeo}}$-RST $\F{T}$ with $\rootterm(\F{T}) = \tgeo(\tz)$ and $\ctheight(\F{T}) = \omega$.
    The node $\textcolor{purple}{v}$ is labeled by the probability $p_{\textcolor{purple}{v}} = \nicefrac{1}{8}$ 
    and the term $a_{\textcolor{purple}{v}} = \ts(\ts(\tz))$, and it has depth $\ctdepth(\textcolor{purple}{v}) = 3$.}\label{fig:RST_example}
    \vspace*{-12px}
\end{figure}

The \emph{termination probability}
of a $\to$-RST $\F{T}$ is
$|\F{T}| = \sum_{v \in \ctleaf(\F{T})} p_v$. 
A PARS $(A, \to)$ is \emph{almost-surely terminating} ($\AST$) if $|\F{T}| = 1$ 
for all $\to$-RSTs $\F{T}$, i.e., the termination probability is 1. 
However, $\AST$ does not imply that 
the expected num\-ber of rewrite steps is finite. 
The \emph{expected derivation length} of
a $\to$-RST $\F{T}$ is $\edl(\F{T}) = \infty$ if $|\F{T}| < 1$,
and $\textstyle \edl(\F{T}) = \sum_{v \in \ctleaf(\F{T})} \ctdepth(v) \cdot p_v$,
otherwise.
For the ${\PP_{\tgeo}}$-RST $\F{T}$ in \cref{fig:RST_example} we get $\edl(\F{T}) =
1 \cdot \nicefrac{1}{2} + 2 \cdot \nicefrac{1}{4} + 3 \cdot \nicefrac{1}{8} + \ldots =
2$,
so in expectation we perform $2$ rewrite steps.
A PARS $(A, \to)$ is \emph{positively almost-surely terminating} ($\PAST$)
if $\edl(\F{T})$ is finite for all $\to$-RSTs $\F{T}$.
These notions  are equivalent to the ones in
\cite{avanzini2020probabilistic,bournez2005proving,kassinggiesl2023iAST}
where $\PASTASTColor$ is defined via a lifting of $\to$
to multisets\linebreak[3] or via stochastic processes.
Note that requiring \emph{fully evaluated} RSTs (where all leaves are normal forms) 
does not change the class of PTRSs that are  $\PASTASTColor$.

\begin{remark}
    When rewriting a term, both the position of the rewrite step and
    the applied rule are chosen \emph{non-deterministically}.
    PTRSs extend this non-determinism by an additional probabilistic choice. 
    Thus, $\PASTASTColor$ for PTRSs corresponds to $\PASTASTColor$
    under all schedulers that resolve the non-probabilistic non-determinism.
\end{remark}

%% file: embedBasics.tex
\section{Disproving $\AST$ and $\PAST$ via Loops and Embeddings}\label{Embedding Random Walks}

The most common approach to disprove termination of a TRS $\R$
is by finding \emph{loops}.
A loop is a sequence $t \to_{\R}^{+} C[t\sigma]$ for some term $t$, context $C$, and substitution $\sigma$. 
It gives rise to an infinite rewrite sequence 
$t \to_{\R}^{n} C[t\sigma] \to_{\R}^{n} C[C[t\sigma]\sigma] \to_{\R} \ldots$
for some $n \in \IN_{>0}$.
However, to disprove $\PASTASTColor$ of a PTRS $\PP$,
a loop in the \emph{non-}\linebreak[3]\emph{probabilistic variant}
$\nonprob(\PP) = \{\ell \to r \mid \ell \to \mu \in \PP, r \in \Supp(\mu)\}$
is not  sufficient.

\begin{example}\label{ex:loop-not-enough}
    Consider the PTRS $\PP_1$ with the rule 
    $\tg(x) \to \{\nicefrac{2}{3}: x, \nicefrac{1}{3}: \tg(\tg(x))\}$ 
    modeling a negatively biased random walk on the number of $\tg$'s.
    Here, $\nonprob(\PP_1)$ contains the two rules $\tg(x) \to x$ and $\tg(x) \to \tg(\tg(x))$.
    The latter gives rise to the loop $\tg(x) \to C[\tg(x)\sigma]$ 
    with $C = \tg(\square)$ and the identity substitution $\sigma$.
    However, $\PP_1$ corresponds to the random walk $\mu_4$ from \Cref{fig:all}, and thus, it is $\PAST$.
\end{example}

Instead of $\nonprob(\PP)$, one has to consider $\PP$-RSTs. If there is a $\PP$-RST with
root $t$ and an instance of  $t$ in every leaf, then $\PP$ is not $\AST$ and thus, \paper{\vspace*{-.4cm}\pagebreak} also not $\PAST$.

\begin{restatable}[Disproving $\PASTASTColor$ via Loops]{theorem}{EmbeddingLoopWalks}\label{theorem:embedding_loop_walks}
  Let $\PP$ be a PTRS and $\F{T}$ be a $\PP$-RST with $\ctheight(\F{T}) > 0$
  where $\rootterm(\F{T}) = t$ and for every $v \in \ctleaf(\F{T})$ 
  there is a context\linebreak[3] $C_v$ and a substitution $\sigma_v$ with $t_v = C_v[t \sigma_v]$.
  Then $\PP$ is neither $\AST$ nor $\PAST$.
\end{restatable}

However, \Cref{theorem:embedding_loop_walks} can only be used to disprove $\PASTASTColor$ for a
very restricted class of PTRSs.  
To handle more complex examples, instead of
requiring that a looping term $t$ occurs in every leaf of a $\PP$-RST,
one has to find a PARS $(A, \to)$ 
and a $\to$-RST $\F{T}$ where it is known that $|\F{T}| < 1$ (to disprove $\AST$)
or $\edl(\F{T}) = \infty$ (to disprove $\PAST$) holds.
In addition, one has to find 
a $\PP$-RST $\F{T}'$ and an \emph{embedding} 
$\mathtt{e} : V(\F{T}) \to V(\F{T}')$ which ensures
$|\F{T}'| \leq |\F{T}|$ and
$\edl(\F{T}) \leq \edl(\F{T}')$.
If $|\F{T}| < 1$, then this disproves $\AST$ of $\PP$, and if
$\edl(\F{T}) = \infty$, then it disproves $\PAST$ of $\PP$.

\begin{definition}[RST-Embedding]\label{def:embedding}
  Let $i \in \{1,2\}$, let $(A_i, \to_i)$ be a PARS, and let $\F{T}_i$ be a $\to_i$-RST. 
  An \emph{embedding} from $\F{T}_1$ to $\F{T}_2$ is an injective mapping 
  $\mathtt{e} : V(\F{T}_1) \to V(\F{T}_2)$ such that
  $p_v = p_{\mathtt{e}(v)}$ for all $v \in V(\F{T}_1)$ (the probability of $v$ in $\F{T}_1$ 
  and the probability of its image $\mathtt{e}(v)$ in $\F{T}_2$ are the same)
  and if  there is a path from a node $v$ to $w$ in $\F{T}_1$,
  then there is a path from $\mathtt{e}(v)$ to $\mathtt{e}(w)$ in $\F{T}_2$ as well.
  Moreover, if $w$ is a direct successor of $v$ in $\F{T}_1$, then
 the path
  from $\mathtt{e}(v)$ to $\mathtt{e}(w)$ in $\F{T}_2$ does not contain any other nodes
  from  the image $\mathtt{e}(V(\F{T}_1))$.
\end{definition}

\vspace*{-.1cm}

\begin{restatable}[Disproving $\PASTASTColor$ via Embeddings]{theorem}{LowerBounds}\label{theorem:lower_bounds_by_embeddings}
  Let ${(A_1, \to_1)}$, ${(A_2, \to_2)}$ be two PARSs, and let $\F{T}_i$ be a $\to_i$-RST for $i \in \{1,2\}$
  such that there exists an embedding from $\F{T}_1$ to $\F{T}_2$.
  Then, $|\F{T}_2| \leq |\F{T}_1|$ and $\edl(\F{T}_1) \leq \edl(\F{T}_2)$.
\end{restatable}

%% file: embedRW.tex
\section{Embedding Random Walks Based on Term Occurrences}\label{Occ Random Walks}

To embed symmetric or even positively biased random walks, 
we again search for a loop where $t$ rewrites to $C[t\sigma]$, but
now we also  consider the \emph{probabilities} and
the \emph{number of loops} represented by the right-hand side $C[t\sigma]$.

\begin{example}\label{ex:three-gs}
    The PTRS $\PP_2$ with the rule 
    $\tg(x) \to \{\nicefrac{2}{3}: x, \nicefrac{1}{3}: \tg(\tg(\tg(x)))\}$ 
    models a symmetric random walk on the number of $\tg$'s,
    because their number increases in each rule application by 
    $\nicefrac{2}{3} \cdot (-1) + \nicefrac{1}{3} \cdot 2 = 0$ in expectation.
    While $\PP_1$ from \Cref{ex:loop-not-enough} is $\PAST$, $\PP_2$ is not.
    The difference is that the symbol $\tg$ occurs three instead of two times 
    in the second choice of the distribution on the right-hand side.
\end{example}

\Cref{ex:three-gs} illustrates that it is important how often 
a looping term like $\tg(x)$ occurs on the right-hand side. 
In this example, the three subterms of $\tg(\tg(\tg(x)))$ that are instantiations of $\tg(x)$
can indeed be counted separately, because they do not overlap at a non-variable position.
Thus, we first solve the problem of finding the maximal number 
of such non-overlapping instantiations (called \emph{occurrences}).

\begin{definition}[Term Occurrences, $\subtermocAlt_\pi$]
    Let $t, s \in \TT$.
    We say that $t$ \defemph{occurs} in $s$ at position $\pi$ 
    (denoted $t \subtermocAlt_{\pi} s$) if $s|_\pi = t \sigma$ for some substitution $\sigma$. 
    Two positions $\pi_1$ and $\pi_2$ are \defemph{overlapping} w.r.t.\  $t$
    if there is a  $\tau \in \pos_{\Sigma}(t)$ with $\pi_1 = \pi_2.\tau$ or
    $\pi_2 = \pi_1.\tau$.
    Let $\NOO(t,s)$ be
    the set of all sets of \defemph{pairwise \textbf{n}on-overlapping \textbf{o}ccurrences} 
    of $t$ in $s$.
    So for $S \subseteq \pos(s)$ we have $S \in \NOO(t,s)$ iff
    $t \subtermocAlt_{\pi} s$ for all $\pi \in S$,\linebreak 
    and $\pi_1, \pi_2 \in S$ with $\pi_1 \neq \pi_2$ implies that $\pi_1, \pi_2$ are non-overlapping w.r.t.\ $t$.
    Let $\maxNOO(t,s) = \max_{S \in \NOO(t,s)} |S|$ be the maximal cardinality of sets in $\NOO(t,s)$.
\end{definition}
So if $t \subtermocAlt_{\pi_1} s$ and $t \subtermocAlt_{\pi_2} s$
where $\pi_1, \pi_2$ are
non-overlapping w.r.t.\ $t$, then $\pi_1, \pi_2$ are 
either orthogonal, or we have $s|_{\pi_1} = t\sigma_1$
and $t\sigma_2 \trianglelefteq \sigma_1(x)$
(or $s|_{\pi_2} = t\sigma_2$
and $t\sigma_1 \trianglelefteq \sigma_2(x)$)
for some variable $x \in \VSet(t)$
and substitutions $\sigma_1, \sigma_2$.

\setlength{\textfloatsep}{5pt}
\begin{algorithm}[btp]
    \scriptsize
    \caption{Compute $\maxNOO(t,s)$ for $t \notin \VSet$}
    \label{alg:maxNOO}
    \DontPrintSemicolon
    $q \gets \texttt{EmptyList()}$\;
    \For{$s' \trianglelefteq s$}{
        $\alpha_{s'} \gets 0$, $\beta_{s'} \gets 0$, $\gamma_{s'} \gets 0$\tcp*[r]{Initialize values}
    }
    \texttt{q.enqueue(leaves)}\tcp*[r]{Start with the leaves}
    \While{\normalfont{\texttt{not q.isEmpty()}}}{
        $s' \gets \texttt{q.dequeue()}$\;
        $\gamma_{s'} \gets 1$\;

        \If{$s'$ \normalfont{\texttt{is leaf}}}{
            $\alpha_{s'} \gets 
            \begin{cases}
                1, & \text{if } t \subtermocAlt_{\varepsilon} s' \hspace*{5.3cm} \textcolor{purple}{\textit{ // } \mathit{t} \textit{ matches } \mathit{s'}}\\
                0, & \text{otherwise} \hspace*{4.4cm} \textcolor{purple}{\textit{ // } \mathit{t} \textit{ does not match } \mathit{s'}}
            \end{cases}$\hspace*{-.5cm}
        }
        \Else{
            $\beta_{s'} \gets \sum_{s' = f(v_1, \dots, v_k)} \sum_{j=1}^{k} \alpha_{v_j}$\;

            $\alpha_{s'} \gets 
            \begin{cases}
                \max\left\{\beta_{s'}, \sum_{\pi \in \pos_{\VSet}(t)} \alpha_{s'|_{\pi}} + 1\right\}, 
                & \text{if } t \subtermocAlt_{\varepsilon} s' \hspace*{1.255cm} \textcolor{purple}{\textit{ // } \mathit{t} \textit{ matches } \mathit{s'}}\\
                \beta_{s'}, & \text{otherwise} \hspace*{.355cm} \textcolor{purple}{\textit{ // } \mathit{t} \textit{ does not match } \mathit{s'}}
            \end{cases}$\hspace*{-.5cm}
        }
        \If{\normalfont{$\forall v \in s'.\texttt{parent().children()}: \gamma_v = 1$}\tcp*[r]{All siblings of $s'$ processed}}{
            \texttt{q.enqueue($s'$.parent())}\tcp*[r]{Enqueue parent}
        }
    }
    \Return $\alpha_s$\; 
\end{algorithm}

For example, we have $\tg(x) \subtermocAlt_{\pi} \tg(\tg(\tg(x)))$ for all $\pi \in \{
\epsilon, 1, 1.1 \}$ and $\NOO(\tg(x),\linebreak[3] \tg(\tg(\tg(x))))$ consists of all subsets of 
$\{\epsilon, 1, 1.1 \}$. Thus,
$\maxNOO(\tg(x), \tg(\tg(\tg(x)))) = 3$. 
So there are three non-overlapping occurrences of the term $\tg(x)$ in $\tg(\tg(\tg(x)))$, 
i.e., these occurrences can indeed all be counted when analyzing non-termination.

On the other hand, $\tg(\tg(x)) \subtermocAlt_{\pi} \tg(\tg(\tg(x)))$ only holds for $\pi \in \{\epsilon, 1 \}$,
and 
$\NOO(\tg(\tg(x)), \tg(\tg(\tg(x)))) = \{ \emptyset, \{\epsilon\}, \{1\}
\}$, but $\{\epsilon, 1 \} \notin \NOO(\tg(\tg(x)), \tg(\tg(\tg(x))))$, because $\epsilon$ and $1$ are
overlapping w.r.t.\ $\tg(\tg(x))$, since $1 \in \pos_{\Sigma}(\tg(\tg(x)))$. Thus, we just 
obtain
$\maxNOO(\tg(\tg(x)), \tg(\tg(\tg(x)))) = 1$, i.e., we can only count one instead of
two occurrences of the term $\tg(\tg(x))$ in $\tg(\tg(\tg(x)))$.
Such overlapping occurrences cannot be counted separately,
because rewriting one occurrence may interfere with 
the possibility to rewrite the other one, see \Cref{ex:non-overlapping-required}.

Computing $\maxNOO(t,s)$ can be done via a
dynamic programming algorithm traversing the tree structure of $s$ from the 
leaves to the root, see \Cref{alg:maxNOO}.
For every node, i.e., every subterm $s' \trianglelefteq s$, we compute
(1) $\alpha_{s'} = \maxNOO(t,s')$
and (2) $\beta_{s'} = \max_{S \in \NOO(t,s'), \varepsilon \notin S} |S|$, 
i.e., the maximal number of non-overlapping occurrences of $t$ in $s'$ if we do not
consider an occurrence of $t$
at the root of $s'$.
For the leaves, we set $\beta_{s'} = 0$ and $\alpha_{s'} = 1$ if $t \subtermocAlt_{\varepsilon} s'$ 
and $\alpha_{s'} = 0$ otherwise. 
For each inner node  where $s'$ has the form
$f(v_1,\ldots,v_k)$,  we set $\beta_{s'} =
\sum_{1 \leq j \leq k} \alpha_{v_j}$,
and then check 
whether there is an occurrence of $t$ at the root of $s'$, i.e.,
whether $t \subtermocAlt_{\epsilon} s'$. 
If not, then we set $\alpha_{s'} = \beta_{s'} =
\sum_{1 \leq j \leq k} \alpha_{v_j}$. 
If there is an occurrence of $t$ at the root of $s'$, then 
$\alpha_{s'} = \max\{\beta_{s'},\sum_{\pi \in \pos_{\VSet(t)}} \alpha_{s'|_{\pi}} + 1\}$,
i.e., $\alpha_{s'}$ is the maximum of
$\beta_{s'}$ and the number obtained when considering the root position and 
$\maxNOO$ for all subterms of $s'$ corresponding to variable positions of $t$.
In \Cref{alg:maxNOO},  we
use a flag $\gamma_{s'}$ where $\gamma_{s'} =1$ indicates that we have already computed
$\alpha_{s'}$ and $\beta_{s'}$, and otherwise we have  $\gamma_{s'} = 0$. \pagebreak[3]
Concerning the runtime,  \Cref{alg:maxNOO} 
has to consider each subterm $s'$ of $s$ and check whether $t$ matches $s'$,
which runs in time $\mathcal{O}(|s'|)$, where $|s'|$ is the number of positions in $s'$. 
So \Cref{alg:maxNOO} runs in $\mathcal{O}(|s|^2)$.

\begin{example}\label{ex:maxnoo}
   \Cref{fig:example_maxNOO} shows the dynamic programming 
 table for 
    $t = \tf(\ta, \tf(\ta,x))$ and $s = \tf(\ta, \tf(\ta, \tf(\ta, \tf(\tf(\ta, \tf(\ta,
 \ta)), \tf(\ta, \tf(\ta,\ta))))))$ computed by \Cref{alg:maxNOO}.
    There is no occurrence of $t$ in $s_{i}$ for any $1 \leq i \leq 11$.
   For $s_{12}$ and $s_{13}$ there is one occurrence of $t$ at the root. 
    Therefore, 
    $\alpha_{s_{12}} = \alpha_{s_{13}} = 1$.
    At $s_{14}$ we have no occurrence of $t$ at the root, but  two occurrences below the root, 
    thus $\alpha_{s_{14}} = \beta_{s_{14}} = 2$.
    The same holds for $s_{15}$.
    We have another occurrence at the root of $s_{16}$    that does not overlap 
    with the occurrences at the roots of $s_{12}$ and $s_{13}$, so $\alpha_{s_{16}} = 3$.
 Finally, there is another occurrence at the root of $s_{17}$,
    but it is overlapping (w.r.t.\ $t$) with the preceding occurrence at the root of $s_{16}$.
    Thus,  $\maxNOO(t,s) = \alpha_{s} = \alpha_{s_{17}} = 3$.
\end{example}

\setlength{\textfloatsep}{\oldtextfloatsep}
\begin{figure}[t]
    \vspace*{-10px}
    \centering
    \begin{minipage}{0.4\textwidth}
        \vspace*{18px}
        \begin{subfigure}{\linewidth}
            \centering
            \begin{tabular}{c@{\hskip 1cm}c@{\hskip 1cm}c}
                \toprule
                Subterm $s'$ & $\alpha_{s'}$ & $\beta_{s'}$ \\
                \midrule
                $s_1$--$s_{11}$ & $0$ & $0$ \\
                $s_{12}$--$s_{13}$ & $1$ & $0$ \\
                $s_{14}$ & $2$ & $2$ \\
                $s_{15}$ & $2$ & $2$ \\
                $s_{16}$ & $3$ & $2$ \\
                $s_{17}$ & $3$ & $3$ \\
                \bottomrule
            \end{tabular}
            \caption{Dynamic Programming Table}
        \end{subfigure}
    \end{minipage}
    \hfill
    \begin{minipage}{0.32\textwidth}
        \begin{subfigure}{\linewidth}
            \centering
            \scriptsize
            \begin{tikzpicture}
                \tikzset{label distance=-2mm}
                \tikzstyle{empty}=[rectangle,thick,minimum size=1mm]
                \node[empty, label=above right:{\tiny \textcolor{purple}{$s_{17}$}}] at (-0.5, 0.5)  (prea) {$\tf$};

                \node[empty, label=above left:{\tiny \textcolor{purple}{$s_{9}$}}] at (-1, 0)  (preb) {$\ta$};
                \node[empty, label=above right:{\tiny \textcolor{purple}{$s_{16}$}}] at (0, 0)  (a) {$\tf$};

                \node[empty, label=above left:{\tiny \textcolor{purple}{$s_{8}$}}] at (-0.5, -0.5)  (b) {$\ta$};
                \node[empty, label=above right:{\tiny \textcolor{purple}{$s_{15}$}}] at (0.5, -0.5)  (c) {$\tf$};

                \node[empty, label=above left:{\tiny \textcolor{purple}{$s_{7}$}}] at (0, -1)  (d) {$\ta$};
                \node[empty, label=above right:{\tiny \textcolor{purple}{$s_{14}$}}] at (1, -1)  (e) {$\tf$};

                \node[empty, label=above left:{\tiny \textcolor{purple}{$s_{13}$}}] at (0.5, -1.5)  (f) {$\tf$};
                \node[empty, label=above right:{\tiny \textcolor{purple}{$s_{12}$}}] at (1.5, -1.5)  (g) {$\tf$};

                \node[empty, label={[label distance=-1mm]below:{\tiny \textcolor{purple}{$s_{6}$}}}] at (0.25, -2)  (i) {$\ta$};
                \node[empty, label={above right:{\tiny \textcolor{purple}{$s_{11}$}}}] at (0.75, -2)  (h) {$\tf$};
                \node[empty, label={[label distance=-1mm]below:{\tiny \textcolor{purple}{$s_{5}$}}}] at (1.25, -2)  (j) {$\ta$};
                \node[empty, label={above right:{\tiny \textcolor{purple}{$s_{10}$}}}] at (1.75, -2)  (k) {$\tf$};

                \node[empty, label={[label distance=-1mm]below:{\tiny \textcolor{purple}{$s_{4}$}}}] at (0.5, -2.5)  (h1) {$\ta$};
                \node[empty, label={[label distance=-1mm]below:{\tiny \textcolor{purple}{$s_{3}$}}}] at (1, -2.5)  (h2) {$\ta$};
                \node[empty, label={[label distance=-1mm]below:{\tiny \textcolor{purple}{$s_{2}$}}}] at (1.5, -2.5)  (k1) {$\ta$};
                \node[empty, label={[label distance=-1mm]below:{\tiny \textcolor{purple}{$s_{1}$}}}] at (2, -2.5)  (k2) {$\ta$};

                \draw (prea) edge[-] (preb);
                \draw (prea) edge[-] (a);
                \draw (a) edge[-] (b);
                \draw (a) edge[-] (c);
                \draw (c) edge[-] (d);
                \draw (c) edge[-] (e);
                \draw (e) edge[-] (f);
                \draw (e) edge[-] (g);
                \draw (f) edge[-] (h);
                \draw (f) edge[-] (i);
                \draw (g) edge[-] (j);
                \draw (g) edge[-] (k);
                \draw (h) edge[-] (h1);
                \draw (h) edge[-] (h2);
                \draw (k) edge[-] (k1);
                \draw (k) edge[-] (k2);
            \end{tikzpicture}
            \caption{Tree Structure of $s$}
        \end{subfigure}
    \end{minipage}
    \hfill
    \begin{minipage}{0.26\textwidth}
        \begin{subfigure}{\linewidth}
            \centering
            \scriptsize
            \begin{tikzpicture}
                \tikzset{label distance=-2mm}
                \tikzstyle{empty}=[rectangle,thick,minimum size=1mm]

                \node[empty] at (0, 0)  (a) {$\tf$};

                \node[empty] at (-0.25, -0.5)  (b) {$\ta$};
                \node[empty] at (0.25, -0.5)  (c) {$\tf$};

                \node[empty] at (0, -1)  (d) {$\ta$};
                \node[empty] at (0.5, -1)  (e) {$x$};

                \node[empty] at (0, 2.5)  (f) {};

                \draw (a) edge[-] (b);
                \draw (a) edge[-] (c);
                \draw (c) edge[-] (d);
                \draw (c) edge[-] (e);
            \end{tikzpicture}
            \caption{Tree Structure of $t$}
        \end{subfigure} 
    \end{minipage}
    \vspace*{-5px}
    \caption{Computation of $\maxNOO(t,s)$}\label{fig:example_maxNOO}
    \vspace{-13px}
\end{figure}

We now embed random walks based on the maximal number of non-overlapping occurrences of a looping term $t$.
We start with a $\PP$-RST $\F{T}$ with $\rootterm(\F{T}) = t$ where $t$ is
\emph{linear} (see \Cref{ex:linearity_required} for the reason).
As usual,  $t \in \TT$ is \emph{linear} if $|t|_x \leq 1$ for all $x \in \VSet$,
where $|t|_x$ is the number of positions $\pi \in \pos(t)$ such that $t|_{\pi} = x$.

Then, we extend the RST $\F{T}$ by rewriting the terms in its leaves,
until  there are enough non-overlapping
occurrences of $t$ in the leaves to disprove $\PASTASTColor$.
In practice, one stops after reaching a suitable finite RST $\F{T}$, but
\Cref{theorem:embedding}
also holds for infinite RSTs $\F{T}$.
We can then use the (finite) tree $\F{T}$ as a representation of an infinite $\PP$-RST $\F{T}^{\infty}$,
where we can embed a random walk that disproves $\PASTASTColor$.

\begin{restatable}[Embedding Random Walks via Occurrences (1)]{theorem}{EmbeddingRW}\label{theorem:embedding}
    Let $\PP$ be a PTRS and let $\F{T}$ be a 
    $\PP$-RST with $\ctheight(\F{T}) > 0$ and
    $\rootterm(\F{T}) = t$, where $t$ is linear. 
    If we have ${\sum_{v \in \ctleaf(\F{T})} p_v \cdot \maxNOO(t,t_v)} > 1$, then $\PP$ is not $\AST$.
    Moreover, if ${\sum_{v \in \ctleaf(\F{T})} p_v \cdot \maxNOO(t,t_v)} \geq 1$, then $\PP$ is not $\PAST$.
\end{restatable}

So for every leaf $v$, we multiply its probability $p_v$ with the number of
non-over\-lapping occurrences of $t$ in the term $t_v$ of the leaf.
This number of occurrences yields  a random walk with $\mu(x) =
\sum_{v \in \ctleaf(\F{T}), x = \maxNOO(t,t_v)-1} p_v$ for all $x \in \IZ$.
By re\-peatedly rewriting an innermost\footnote{The innermost strategy must be used due to
rules where a variable occurs more often on the left-
than on the right-hand side. Such rules
might decrease the number of occurrences of $t$ in the arguments when
rewriting at a non-innermost position.}
occurrence of $t$ in the leaves of the tree accord\-ing to the rules used in $\F{T}$,
we obtain an infinite $\PP$-RST
$\F{T}^{\infty}$, where we can embed the computation of the random walk $\mu$.
This  yields the desired lower bound.
Here, it is not a problem if different instantiations of $t$ occur at different
positions in\linebreak[3] a leaf.
The reason is that if $t$ starts a loop, then so does every instantiation of $t$.

\begin{example}[Embedding via Occurrences and Innermost Rewriting]\label{ex:embedding_innermost}
    Consider the PTRS $\PP_3$ with the rules $\tg(x) \to \{\nicefrac{1}{2}:x, \nicefrac{1}{2}:\tf(x)\}$ 
    and $\tf(x) \to \{1:\tg(\tg(x))\}$.
    If we count the occurrences of $\tg(x)$, 
    then the $\PP_3$-RST $\F{T}$ gives rise to the symmetric
    random walk $\mu$ both depicted in \Cref{fig:embedding_one},
    which represents a ``lower bound'' for the termination behavior of $\F{T}$.
Here, we represent $\mu$ as a PARS $(\IZ, \hookrule_{\mu})$ with 
    $y \hookrule_{\mu} \{(p:y+\maxNOO(t,t_v)-1) \mid v \in \ctleaf(\F{T})\}$ for every $y
> 0$. The $\hookrule_{\mu}$-RST with infinite expected derivation length
can be embedded in the tree that results from extending
$\F{T}$ by rewriting the innermost subterm $\tg(x)$ repeatedly according to the rules used in $\F{T}$. 
So since $\mu$ is not $\PAST$, $\PP_3$ is not $\PAST$ either.
    However, since $\mu$ is $\AST$,
    this does not yield any information about whether $\PP_3$ is $\AST$.
\end{example}

\vspace*{-.3cm}

\begin{figure}[t]
    \vspace{-2px}
    \begin{center}
        \scriptsize
        \begin{tikzpicture}
        \tikzstyle{adam}=[thick,draw=black!100,fill=white!100,minimum size=4mm,
            shape=rectangle split, rectangle split parts=2,rectangle split horizontal,font={\scriptsize}]
        \tikzstyle{empty}=[rectangle,thick,minimum size=4mm]

        \node[empty] at (1, 0.6)  (header1) {\small $\PP_3$-RST $\F{T}$};
        \node[adam] at (1, 0)  (a) {$1$\nodepart{two}$\tg(x)$};

        \node[adam] at (0, -0.7)  (b) {$\nicefrac{1}{2}$\nodepart{two}$x$};
        \node[adam] at (2, -0.7)  (c) {$\nicefrac{1}{2}$\nodepart{two}$\tf(x)$};

        \node[adam] at (2, -1.4)  (d) {$\nicefrac{1}{2}$\nodepart{two}$\tg(\tg(x))$};

        \node[empty] at (2, -2.1)  (dend) {$\ldots$};

        \node[empty] at (6, 0.6)  (header2) {\small $\hookrule_{\mu}$-RST};
        \node[adam] at (6, 0)  (a2) {$1$\nodepart{two}$1$};

        \node[adam] at (5, -0.7)  (b2) {$\nicefrac{1}{2}$\nodepart{two}$0$};
        \node[adam] at (7, -0.7)  (c2) {$\nicefrac{1}{2}$\nodepart{two}$2$};

        \node[empty] at (7, -1.4)  (c2end) {$\ldots$};
        
        \draw (a) edge[->] (b);
        \draw (a) edge[->] (c);
        \draw (c) edge[->] (d);
        \draw (d) edge[->] (dend);
        \draw (a2) edge[->] (b2);
        \draw (a2) edge[->] (c2);
        \draw (c2) edge[->] (c2end);

        \draw[dashed, ->, bend right=15] (a2) to node [empty, above] {\footnotesize $\mathtt{e}$} (a);
        \draw[dashed, ->, bend right=20] (b2) to (b);
        \draw[dashed, ->, bend left=10] (c2) to (d);

        \node[empty] at (9.5, 0.6)  (header2) {\small Random Walk $\mu$};
        \draw (9.5,-1.35) node{$\begin{aligned}
        \mu(-1) =\;& \nicefrac{1}{2} \\
        \mu(1) =\;& \nicefrac{1}{2} \end{aligned}$};
        
        \draw (8.5,-0.6) --++(90:1cm);
        \fill[pattern=north west lines] (8.5,-0.6) rectangle ++(-0.5,1); 

        \draw[->, thick] (8,-0.6) -- (11.5,-0.6);
        \foreach \x/\den in {
            0/50, 1/50, 2/50} {
            \node[circle, fill=gray!\den, inner sep=0pt, minimum size=0.35cm] at (\x+8.5,-0.6) {\x};
        }

        \draw[-Stealth, dashed, thick, yshift=3pt, shorten >= 2pt, shorten <= 2.5pt] (9.5,-0.6) to[bend left] node[above right] {$\nicefrac{1}{2}$} (10.5,-0.6);
        \draw[-Stealth, dashed, thick, yshift=3pt, shorten >= 2pt, shorten <= 2.5pt] (9.5,-0.6) to[bend right] node[above left] {$\nicefrac{1}{2}$} (8.5,-0.6);
        \end{tikzpicture}
    \end{center}
    \vspace{-15px}
    \caption{Embedding of a Random Walk into a $\PP_3$-RST to Disprove $\PAST$ of $\PP_3$}\label{fig:embedding_one}
    \vspace{-11px}
\end{figure}

\begin{example}[Non-Overlappingness is Required]\label{ex:non-overlapping-required}
    Non-overlappingness of the different occurrences of $t$ in a leaf guarantees that rewriting an innermost
occurrence of $t$
does not interfere with the possibility to rewrite the other occurrences\linebreak of $t$ later on.
To see this,
    consider the PTRS $\PP_4$ with the rule
    $\tg(\tg(x)) \to \{\nicefrac{1}{3}: x, \nicefrac{2}{3}: \tg(\tg(\tg(x)))\}$, 
    which is $\AST$.
    If one counted both occurrences of $\tg(\tg(x))$ in the term $\tg(\tg(\tg(x)))$ in spite
    of their overlap, then one could embed the random walk $\mu_3$ from \Cref{fig:all}, 
    and thus, falsely disprove $\AST$ of $\PP_4$. Here, the problem is that rewriting the
    innermost subterm $\tg(\tg(x))$ of $\tg(\tg(\tg(x)))$ could yield $\tg(x)$, i.e., then
    the outermost occurrence  of $\tg(\tg(x))$ in $\tg(\tg(\tg(x)))$ would be ``destroyed''.
\end{example}

\begin{example}[Linearity of $t$ is Required]\label{ex:linearity_required}
    Linearity of $t$ is required in \Cref{theorem:embedding}, because otherwise rewriting
    an 
    innermost occurrence of $t$
    in a leaf may ``destroy'' other occurrences of $t$ in that leaf.
    For example, 
    consider the PTRS $\PP_5$ with the rule 
    $\tf(x,x) \to \{\nicefrac{1}{3}: \ta, \nicefrac{1}{3}: \tb,
    \nicefrac{1}{3}: \tf(\tf(x,x), \tf(x,x))\}$.
    If we count the occurrences of  $\tf(x,x)$, 
    then the $\PP_5$-RST $\F{T}$ where we perform a single rewrite step starting in $\tf(x,x)$ 
    gives rise to the random walk $\mu_5$ with $\mu_5(-1) = \nicefrac{2}{3}$ and $\mu_5(2)
    = \nicefrac{1}{3}$, since $\maxNOO(\tf(x,x), \tf(\tf(x,x), \tf(x,x))) = 3$.
    Since $\mu_5$ is not $\PAST$, we would falsely disprove $\PAST$ of $\PP_5$.
    But in fact, $\PP_5$ is $\PAST$. The problem is that
    rewriting the proper subterms of $\tf(\tf(x,x), \tf(x,x))$ may yield terms like
    $\tf(\ta,\tb)$,
    where the two
    arguments of $\tf$ are not equal. Thus, rewriting an innermost occurrence of $t
    = \tf(x,x)$ in $\tf(\tf(x,x), \tf(x,x))$
    may ``destroy'' the occurrence of $t$ at the root. 
\end{example}

So when rewriting innermost occurrences of $t$ according to the rules used in $\F{T}$, 
we need linearity of $t$. \pagebreak[3]
Instead, one could also rewrite outermost occurrences.
Then, instead of linearity, we have to require that there is no $v \in \ctleaf(\F{T})$
where a variable occurs more often in the looping term $t$ than in the term $t_v$ of the leaf $v$.
A $\PP$-RST $\F{T}$ is \emph{\textbf{n}on-\textbf{v}ariable-\textbf{d}ecreasing} ($\nvd{}$) if
$|\rootterm(\F{T})|_x \leq |t_v|_x$ for all $v \in \ctleaf(\F{T})$ and all $x \in \VSet$.
Rewriting an outermost occurrence of $t$ according to an $\nvd{}$ $\PP$-RST $\F{T}$ 
does not affect any other non-overlapping occurrences of $t$.

\begin{restatable}[Embedding Random Walks via Occurrences (2)]{theorem}{EmbeddingRWTwo}\label{theorem:embedding_two}
    Let $\PP$ be a PTRS and let $\F{T}$ be an $\nvd{}$
    $\PP$-RST with $\ctheight(\F{T}) > 0$ and $\rootterm(\F{T}) = t$. 
    If we have ${\sum_{v \in \ctleaf(\F{T})} p_v \cdot \maxNOO(t,t_v)} \geqMaybe 1$, 
    then $\PP$ is not $\PASTASTColor$.
\end{restatable}

\begin{example}[Embedding and  Outermost Rewriting]\label{ex:embedding_outermost}
   The PTRS $\PP_5'$ (similar to $\PP_5$ from \Cref{ex:linearity_required})
    has the rule 
    $\tf(x,x) \to \{\nicefrac{1}{3}: \ta(x,x),  \nicefrac{1}{3}: \tb(x,x),
    \nicefrac{1}{3}: \tf(\tf(x,x),\linebreak[3] \tf(x,x))\}$.
    Rewriting at the outermost position yields the $\nvd{}$ $\PP_5'$-RST $\F{T}$ in \Cref{fig:pp5-rst},  
    where $t = \tf(x,x)$. As
    $\maxNOO(t, \ta(t,t)) = \maxNOO(t, \tb(t,t)) = 2$,
    $\maxNOO(t,\linebreak \tf(\tf(t,t), \tf(t, t))) = 7$, and
    $\maxNOO(t, \ta(x,x)) = \maxNOO(t, \tb(x,x)) = 0$, we get\linebreak[3]
    ${\sum_{v \in \ctleaf(\F{T})} p_v \cdot \maxNOO(t,t_v)} = \nicefrac{11}{9} > 1$. So
by \Cref{theorem:embedding_two} this
    disproves $\AST$ of $\PP_5'$.
 \end{example}

\begin{figure}[t]
    \centering
    \begin{minipage}{0.45\textwidth}
        \centering
        \begin{tikzpicture}
        \tikzstyle{adam}=[thick,draw=black!100,fill=white!100,minimum size=4mm,
            shape=rectangle split, rectangle split parts=2,rectangle split horizontal,font={\footnotesize}]
        \tikzstyle{empty}=[rectangle,thick,minimum size=4mm]
    \node[adam] at (0, 0)  (a) {$1$\nodepart{two}$t$};

    \node[adam] at (0, -0.9)  (b) {$\nicefrac{1}{3}$\nodepart{two}$\tf(t,t)$};
   \node[adam] at (-2, -0.9)  (c) {$\nicefrac{1}{3}$\nodepart{two}$\ta(x,x)$};
        \node[adam] at (2, -0.9)  (d) {$\nicefrac{1}{3}$\nodepart{two}$\tb(x,x)$};

        \node[adam] at (0, -1.8)  (e) {$\nicefrac{1}{9}$\nodepart{two}$\tf(\tf(t,t), \tf(t,t))$};
 \node[adam] at (-2.5, -1.8)  (f) {$\nicefrac{1}{9}$\nodepart{two}$\ta(t,t)$};
        \node[adam] at (2.5, -1.8)  (g) {$\nicefrac{1}{9}$\nodepart{two}$\tb(t,t)$};

        \draw (a) edge[->] (b);
        \draw (a) edge[->] (c);
        \draw (a) edge[->] (d);
        \draw (b) edge[->] (e);
      \draw (b) edge[->] (f);
      \draw (b) edge[->] (g);
      \end{tikzpicture}\vspace*{-3px}
        \caption{$\PP_5'$-RST}\label{fig:pp5-rst}
    \end{minipage}
    \hfill
    \begin{minipage}{0.45\textwidth}
        \centering
        \begin{tikzpicture}
                \tikzstyle{adam}=[thick,draw=black!100,fill=white!100,minimum size=4mm,
                shape=rectangle split, rectangle split parts=2,rectangle split horizontal,font={\footnotesize}]
                \tikzstyle{empty}=[rectangle,thick,minimum size=4mm]
                \node[adam] at (0, 0)  (a) {$1$\nodepart{two}$\tf(\tg(x))$};
                \node[adam] at (-1, -0.9)  (b) {$\nicefrac{1}{3}$\nodepart{two}$\tf(x)$};
                \node[adam] at (1, -0.9)  (c) {$\nicefrac{2}{3}$\nodepart{two}$\tf(\tg(\tg(x)))$};
                \node[empty] at (1, 0.9)  (d) {};
            
            \draw (a) edge[->] (b);
            \draw (a) edge[->] (c);
        \end{tikzpicture}\vspace*{-3px}
        \caption{$\PP_6$-RST}\label{fig:pp6-rst}
    \end{minipage}
    \vspace{-12px}
\end{figure}

If $t$ is not linear and the RST $\F{T}$ is not $\nvd{}$, then we can only count \emph{\textbf{o}rthogonal \textbf{o}ccurrences}.
The maximal number of orthogonal occurrences of $t$ in $s$ is $\maxParaNOO(t,s) = \max\{|S| \mid S \in \NOO(t,s), \forall \pi_1, \pi_2 \in S \text{
  with }\pi_1 \neq \pi_2: \pi_1 \bot \pi_2\}$.

To compute $\maxParaNOO(t,s)$, we can adjust \Cref{alg:maxNOO} at Line 12 in the case of $t \subtermocAlt_{\varepsilon} s'$. 
Instead of setting $\alpha_{s'}$ to the maximum of $\beta_{s'}$ and $\sum_{\pi \in \pos_{\VSet}(t)} \alpha_{s'|_{\pi}} + 1$,
we  set $\alpha_{s'}$ to $\max\{\beta_{s'}, 1\}$, because now we do not count occurrences
below another occurrence anymore.
The runtime of the adjusted algorithm is still in $\mathcal{O}(|s|^2)$.

\begin{restatable}[Embedding Random Walks via Occurrences (3)]{theorem}{EmbeddingRWThree}\label{theorem:embedding_three}
  Let $\PP$ be a PTRS  and let $\F{T}$ be a $\PP$-RST with $\ctheight(\F{T}) > 0$ and
  $\rootterm(\F{T}) = t$.
    If we have ${\sum_{v \in \ctleaf(\F{T})} p_v \cdot \maxParaNOO(t,t_v)} \geqMaybe 1$,
    then $\PP$ \pagebreak[3] is not $\PASTASTColor$.
\end{restatable}

To automate \Cref{theorem:embedding}, \ref{theorem:embedding_two}, and \ref{theorem:embedding_three}, 
we have to find a $\PP$-RST satisfying one of the two constraints.
To this end, we first search for a looping term $t$.
Here, we reuse\linebreak the non-probabilistic loop detection algorithms from \cite{Frocos05} on the
non-probabilistic variant $\nonprob(\PP)$. 
After finding a loop of $\nonprob(\PP)$ starting with a term $t$ 
(which is already an undecidable problem in general),
we first check whether $t$ is linear.
If $t$ is linear, then we reconstruct the corresponding $\PP$-RST $\F{T}$ for this path.
On all remaining terms $u$ in the leaves of $\F{T}$, we check whether 
we can possibly reach a term $s$ from $u$ such that  $t
\subtermocAlt_\pi s$ for some $\pi \in \pos(s)$.
This is undecidable in general, but we use the \emph{symbol transition graph} from 
\cite{sternagelReachabilityAnalysisTermination2019}
as a sufficient criterion.
On those terms $u$ where we may potentially reach such a term $s$, 
we extend the RST by performing \pagebreak[3] further \emph{rewrite steps}, 
to obtain leaves that contain more occurrences of $t$.
This is performed repeatedly
until we have constructed an RST that satisfies one of the
conditions of \Cref{theorem:embedding}
or until we reach a threshold for the number of rewrite steps. 
In the latter case, we try to find another looping term to generate an
RST in order to
embed a suitable random walk.
If the looping term $t$ is not linear,
then we proceed in an analogous way to apply
\Cref{theorem:embedding_two} or 
\Cref{theorem:embedding_three}, depending on whether the tree 
is $\nvd{}$ or not. Here,
we compute both $\maxNOO(t, t_v)$ and $\maxParaNOO(t, t_v)$ for every leaf
$v$, since by rewriting the leaves, it may change whether the tree is  $\nvd{}$ or not.

%% file: embednonloops.tex
\vspace*{-.25cm}

\section{Embedding Random Walks Based on Pattern Terms}\label{Pattern Random Walks}

\vspace*{-.05cm}

Instead of counting the occurrences of a single term,
we can also count the number of \emph{instantiations} applied to
a certain \emph{base term}.
To simplify the presentation, we only consider orthogonal occurrences in this section.
Similar to \Cref{Occ Random Walks}, corresponding approaches
that consider pairwise
non-overlapping occurrences and use 
innermost or outermost rewriting
are possible as well.
However, how to automate such improvements is open and an interesting direction for future work.

\vspace*{-.1cm}

\begin{example}\label{ex:occ-not-enough}
    Consider $\PP_6 = \left\{\tf(\tg(x)) \to \{\nicefrac{1}{3}: \tf(x), \nicefrac{2}{3}: \tf(\tg(\tg(x)))\}\right\}$ 
    modeling a positively biased random walk on 
    the number of $\tg$'s \emph{directly below an $\tf$} in a term
    and the corresponding
    $\PP_6$-RST in \Cref{fig:pp6-rst}.
    Here, it is not enough to count the occurrences of $\tf(\tg(x))$ to disprove $\AST$,
    but instead we have to count the occurrences of $\tg(x)$ below an $\tf$ symbol.
    Moreover, the $\tf$ in $\tf(x)$ is crucial.
    The PTRS $\PP_6'$ with the rule 
    $\tf(\tg(x)) \to \{\nicefrac{1}{3}: x, \nicefrac{2}{3}: \tf(\tg(\tg(x)))\}$ 
    does not model a random walk anymore, since for any term $\tf(\tg^n(x))$
    we can directly stop and rewrite to $\tg^{n-1}(x)$ 
    (removing the outer $\tf$). So $\PP_6$ is not $\AST$, but $\PP_6'$ is even $\PAST$.
\end{example}

\vspace*{-.1cm}

We formalize this with the concept of a \emph{pattern term} $\langle t, \sigma \rangle$ that represents all terms which result from applying the
\emph{pumping substitution} $\sigma$ repeatedly to the \emph{base term} $t$, i.e., $\langle t, \sigma \rangle$ represents the terms $t$, $t \sigma$, $t \sigma^2$, etc. 
So if $t = \tf(x)$ and\linebreak[3] $\sigma = [x/\tg(x)]$, 
then $\langle t, \sigma \rangle$ represents the terms $\tf(x)$, $\tf(\tg(x))$, $\tf(\tg(\tg(x)))$, etc.
A similar notion was defined in \cite{emmes2012ProvingNonloopingNontermination} to disprove
\emph{non-looping non-termination}. In contrast,\linebreak
here we want to \emph{count loops}, and we want to count how
often the substitution $\sigma$ is applied.
Therefore, we have to distinguish $t \sigma^{m}$ and $t \sigma^{m'}$ whenever $m \neq m'$.

\vspace*{-.1cm}

\begin{definition}[Pattern Term]\label{def:pattern-term}
    A pair $\langle t, \sigma \rangle$ is a \defemph{pattern term} if $t \sigma^{m} \neq t
    \sigma^{m'}$ holds whenever $m \neq m'$.
    We call $t$ the \defemph{base term} and $\sigma$ the \defemph{pumping substitution}.
\end{definition}

\vspace*{-.1cm}

\noindent
To check whether $\langle t, \sigma \rangle$ is a pattern term,
we use the \emph{variable transition graph}.

\vspace*{-.1cm}

\begin{definition}[Variable Transition Graph]
    For any substitution $\sigma$, the graph
 $G_{\sigma}$ has all variables  as nodes 
    and there is an edge from  $x$
    to $y$ if $y \in \VSet(x \sigma)$.
For any term $t$, 
the \defemph{variable transition graph} of $\sigma$ w.r.t.\  $t$ 
    is the subgraph $G_{\sigma, t}$ of $G_{\sigma}$ which contains only those nodes  
   that are reachable from a node in $\VSet(t)$.
\end{definition}

\vspace*{-.1cm}

 \begin{wrapfigure}[3]{r}{0.12\textwidth}
    \scriptsize
    \vspace*{-.8cm}
    \hspace*{.0cm}\begin{tikzpicture}
        \tikzset{label distance=-2mm}

        \node[circle, draw]  at (0, 0)  (a) {$x$};

        \node[circle, draw] at (0.9, 0)  (b) {$y$};
        \node[rectangle, draw] at (0, -0.7)  (c) {$z$};

        \draw (a) edge[->, bend left] (b);
        \draw (b) edge[->, bend left] (a);
        \draw (a) edge[->] (c);
        \draw (c) edge[->, loop right] (c);
    \end{tikzpicture}  
\end{wrapfigure}
\refstepcounter{theorem}
\noindent \textit{Example~\thetheorem.}
Let $t = \tf(x,y)$ and $\sigma = [x / \tf(z, y), y / x]$.
Then $G_{\sigma, t}$ is shown on
\noindent  the right.
There is an edge from $z$ to $z$ because $z \notin \dom(\sigma)$, and hence, $z \sigma = z$.
Nodes from $\VSet(t) = \{x, y\}$ are marked by  circles.

\medskip

\noindent
$G_{\sigma, t}$ yields a computable, sound, and complete criterion for \paper{\vspace*{-.2cm}\pagebreak}
pattern terms.

\begin{restatable}[Detecting Pattern Terms]{lemma}{SuffAndNecForInequal}\label{lemma:suff_and_nec_for_inequality}
  For a term $t$  and a substitution $\sigma$,\linebreak $\langle t, \sigma \rangle$ is a
  pattern term iff 
  some cycle of 
  $G_{\sigma, t}$ contains a variable $x$ with $x\sigma \notin \VSet$.
  \end{restatable}

\begin{example}
    The pair $\langle t, \sigma \rangle$ with $t = \tf(x,y)$ and $\sigma = [x / \tf(z, y), y / x]$
    is a pattern term, because
$G_{\sigma, t}$ has a cycle that contains the variable $x$ and $x\sigma =  \tf(z, y) 
 \notin \VSet$.
\end{example}

\noindent
Next, we consider the problem of counting orthogonal pattern term occurrences.

\begin{definition}[Pattern Occurrences]\label{def:Pattern Occurrences}
    Let $\langle t, \sigma \rangle$ be a pattern term and let $s \in \TT$ be a term.
    We say that $\langle t, \sigma \rangle$ \defemph{occurs with multiplicity} 
    $m_{\pi} \in \IN$ at position $\pi$ in $s$
    (denoted by $\langle t, \sigma \rangle \subtermocAlt^{m_{\pi}}_{\pi} s$)
    if there is an occurrence $t \subtermocAlt_{\pi} s$
    and $m_{\pi}$ is the maximal number such that $t \sigma^{m_{\pi}} \subtermocAlt_{\pi} s$.
    For every set $S \in \NOO(t,s)$, let $m_S = \sum_{\pi \in S} m_\pi$.
    Let $\maxParaNOM(t, \sigma, s) = \max\{m_S \mid S \in \NOO(t,  s),
\forall \pi_1, \pi_2 \in S \text{
  with }\pi_1 \neq \pi_2: \pi_1 \bot \pi_2\}$
    denote the maximal value that one can obtain
    by adding all multiplicities for a set $S$ of
    pairwise orthogonal occurrences of $\langle t, \sigma \rangle$ in $s$.
\end{definition}

\begin{example}
    As an example, consider the term $t = \tf(x)$, the substitution $\sigma = [x/\tg(x)]$, 
    and $s = \tc(\tf(\tg(x)), \tf(\tg(\tg(x))))$. 
    Then $\langle t, \sigma \rangle$ occurs in $s$ at position $1$ with multiplicity 1,
    and at position $2$ with multiplicity 2. 
    Thus, we have $m_{1} = 1$ and $m_2 = 2$. 
    Since  $\NOO(t,  s) = \{  \emptyset, \{ 1 \}, \{ 2 \}, \{ 1, 2\}\}$ and
    both occurrences are at orthogonal positions,
     we have $\maxParaNOM(t, \sigma, s) = m_{\{ 1, 2\}} = m_{1} + m_{2} = 1 + 2 = 3$.
\end{example}

To compute $\maxParaNOM(t, \sigma, s)$, we
use \Cref{alg:maxNOO} with $t \sigma$ instead of $t$. 
Moreover, we adjust \Cref{alg:maxNOO} at Line 12 if $t\sigma \subtermocAlt_{\varepsilon} s'$. 
Instead of setting $\alpha_{s'}$ to $\max \{ \beta_{s'}, 1\}$
as in the computation of 
$\maxParaNOO(t\sigma,s)$,
we set  $\alpha_{s'}$ to $\max \{ \beta_{s'}, m\}$
where $m$ is the multiplicity of the occurrence of $\langle t, \sigma \rangle$ at the root of $s'$.
Note that we only consider orthogonal occurrences here. So if we 
consider the occurrence at the root of $s'$, then occurrences
below the root are ignored.
To find $m$, we check for occurrences of $t \sigma$, $t \sigma^2$, $\ldots$, $t \sigma^{|s|}$.
Compared to \Cref{alg:maxNOO}, we  perform at most $|s| - 1$ additional 
matching checks in each iteration, resulting in a runtime of $\mathcal{O}(|s|^3)$.

\begin{example}\label{ex:maxnom}
    Consider the pattern term $\langle t', \sigma \rangle$ with base term $t' = \tf(\ta,x)$ 
    and pumping substitution $\sigma = [x / \tf(\ta,x)]$, 
    and let $t = \tf(\ta, \tf(\ta,x))$ and $s$ be as in \Cref{ex:maxnoo}.
    Thus, $t = \tf(\ta,x) [x / \tf(\ta,x)]$.
    We have
    occurrences of $\langle t', \sigma \rangle$ with
    multiplicity 0 at $s_{10}$, $s_{11}$,  $s_{15}$;
    with
    multiplicity 1 at $s_{12}$, $s_{13}$,  $s_{16}$;
    and with 
     multiplicity $2$ 
     at the root $s_{17}$.
Since we only consider occurrences at orthogonal positions, the maximum is obtained by
adding the multiplicities for the orthogonal subterms  $s_{12}$ and $s_{13}$ or by
considering the multiplicity at the root. Thus, 
 $\maxParaNOM(t',\sigma,s) = \alpha_{s} = \alpha_{s_{17}} = 2$.
\end{example}

As demonstrated by the PTRS $\PP_6'$ in \Cref{ex:occ-not-enough}, if
a term $t \sigma^m$ can be rewritten to a term without any occurrence of $t$, 
then this does not mean that the multiplicity is reduced by $m$, 
but it may mean that one directly reaches a normal form. 
Hence, then this pattern cannot be used to disprove $\PASTASTColor$.
So a pattern $\langle t, \sigma \rangle$ may only be used for disproving
$\PASTASTColor$ if we have a $\PP$-RST where every leaf
contains an occurrence of $t$.
Moreover, if we have a $\PP$-RST $\F{T}$ that starts with $t \sigma^1$
and has an occurrence
$\langle t, \sigma \rangle \subtermocAlt^{q}_\pi t_v$ in a leaf $v$,
then we also need that the tree can be ``generalized'' from $1$ to an arbitrary multiplicity $m \in \IN_{>0}$,
i.e., rewriting the term $t \sigma^{m}$ using the same rules as in $\F{T}$ at the same positions
must result in a leaf $v'$ with an occurrence 
$\langle t, \sigma \rangle \subtermocAlt^{q + m - 1}_\pi t_{v'}$.
 For any occurrence $\langle t, \sigma \rangle \subtermocAlt^{q}_\pi t_v$, let
$\kappa_{v, \pi}$ be the substitution such that $t \sigma^q \kappa_{v, \pi} = t_v|_{\pi}$. \pagebreak[3]
Then we ensure this generalization property by requiring commutation of $\sigma$ with all
these
substitutions $\kappa_{v, \pi}$.

\begin{definition}[Pattern Tree]\label{def:pattern-tree}
    Let $\PP$ be a PTRS, let $\F{T}$ be a $\PP$-RST, and let $\langle t, \sigma \rangle$ be a pattern term.
    $\F{T}$ is a $\PP$-\defemph{pattern tree} for $\langle t, \sigma \rangle$
    if $\rootterm(\F{T}) = t \sigma$, for
    every leaf $v \in \ctleaf(\F{T})$ there exists an occurrence $t \subtermocAlt_\pi t_v$,
    and
whenever 
$t \sigma^q \kappa_{v, \pi} = t_v|_{\pi}$ for some $q \geq 0$, some $\pi \in \pos(t_v)$,
and some substitution $\kappa_{v, \pi}$, then 
    the pumping substitution $\sigma$ commutes with  $\kappa_{v, \pi}$,
    i.e., $\sigma \kappa_{v, \pi} = \kappa_{v, \pi} \sigma$.
\end{definition}

For \Cref{ex:occ-not-enough}, the $\PP_6$-RST in \Cref{fig:pp6-rst} is a pattern tree
for the pattern term  $\langle t, \sigma \rangle$ where $t = \tf(x)$ and $\sigma = [x/\tg(x)]$. 
Here,  the substitutions $\kappa_{v_1, \varepsilon}$ and $\kappa_{v_2, \varepsilon}$
for the leaves are the identity, which
commutes with $\sigma$. 
Indeed, this tree can be ``generalized'' to arbitrary multiplicities 
$m>0$, because the corresponding $\PP_6$-RST
starting with $\tf(\tg^m(x))$ at the root has the leaves
$\tf(\tg^{m-1}(x))$ and $\tf(\tg^{2+m-1}(x))$.

Similar to \Cref{theorem:embedding_three},
by rewriting orthogonal occurrences,
we
result in a technique to embed random walks via patterns in a $\PP$-RST.

\begin{restatable}[Embedding RWs via Patterns]{theorem}{EmbeddingRWPattern}\label{theorem:embedding-pattern}
    Let $\PP$ be a PTRS, $\langle t, \sigma \rangle$ be a pattern term, 
    and let $\F{T}$ be a $\PP$-pattern tree for $\langle t, \sigma \rangle$ with $\ctheight(\F{T}) > 0$.
    If we have $\sum_{v \in \ctleaf(\F{T})} p_v \cdot \maxParaNOM(t, \sigma, t_v) \geqMaybe 1$, 
    then $\PP$ is not $\PASTASTColor$.
\end{restatable}

\begin{example}
   By \Cref{theorem:embedding-pattern}, we  embed the positively biased random
    walk $\mu_3$ of Fig.\linebreak \ref{fig:all} with $\mu_3(-1)=\nicefrac{1}{3}$ and  $\mu_3(1)=\nicefrac{2}{3}$ in the
    $\PP_6$-RST of \Cref{fig:pp6-rst} and disprove $\AST$.
\end{example}

\begin{example}
    By counting the number of applications of the pumping substitution $\sigma = [x/\tg(y), y/\tf(x)]$ to the base term $t = \tc(y,x)$,
    we can disprove $\AST$ of the PTRS $\PP_7$ with the only rule 
    $\tc(\tf(x), \tg(y)) \to \{\nicefrac{1}{3}: \tc(y, x), \nicefrac{2}{3}: \tc(\tf(\tg(y)), \tg(\tf(x)))\}$ 
    via \Cref{theorem:embedding-pattern}. Again,
we can embed the random walk $\mu_3$ in the RST corresponding to the only rewrite rule,
    because the child $t = \tc(y, x)$ has the probability $\nicefrac{1}{3}$, 
    and the second child $t \sigma^2 = \tc(\tf(\tg(y)), \tg(\tf(x)))$ has the probability $\nicefrac{2}{3}$.
\end{example}

\begin{remark}
   \Cref{theorem:embedding-pattern} is \emph{not} a generalization of \Cref{theorem:embedding_three}.
    We can disprove $\PAST$ of $\PP_8$ with the rule
    $\tg \to \{\nicefrac{1}{2}:\ta,\nicefrac{1}{2}:\tc(\tg,\tg)\}$ 
    via \Cref{theorem:embedding_three} by  the tree $\F{T}$ 
    corresponding to the only rule and counting the occurrences of $\tg$.
    However, there is no pattern term $\langle t, \sigma \rangle$ with
    $t \sigma = \tg$.
    Indeed, $\PAST$ of $\PP_8$ cannot be disproven
    via \Cref{theorem:embedding-pattern}.
\end{remark}

To automate \Cref{theorem:embedding-pattern}, 
we adapt our implementation of \Cref{theorem:embedding_three}.
After finding the pattern $t \sigma$,
we have to rewrite the leaves until we obtain a pattern tree.
Note that we can directly stop if our reachability analysis 
shows that some leaf cannot reach a term containing an occurrence of $t$.

%% file: conclusion.tex
\section{Evaluation and Conclusion}\label{Conclusion}

We presented the first techniques to disprove $\PASTASTColor$ of PTRSs automatically. 
To this end, we embed random walks in RSTs, based on counting
occurrences of terms or multiplicities of patterns. 
In this way, qualitative approaches to detect
non-termination of non-probabilistic TRSs based on loops or pattern terms can be
lifted to a quantitative analysis
in the probabilistic setting.
Our approach can be based on any algorithm to detect loops for standard non-probabilistic
TRSs.

We implemented our new contributions 
in our termination prover \tool{AProVE}~\cite{JAR-AProVE2017}.
Currently, \pagebreak[3]
we run our techniques to prove and to disprove termination of PTRSs
in parallel and stop once one of the techniques succeeds.
For proving termination, we use the probabilistic \emph{dependency pair} (DP) framework 
\cite{kassing2025DependencyPairsExpecteda,kassing2026AnnotatedDependencyPaira}, 
which allows to apply different techniques to different sub-problems.
In the future, we plan to integrate our new techniques to disprove
 $\PASTASTColor$
into the DP frameworks for $\AST$ \cite{kassing2026AnnotatedDependencyPaira} 
and $\PAST$ \cite{kassing2025DependencyPairsExpecteda}, respectively.
Then the DP framework can help to
restrict the search for non-termination proofs to
those parts of a PTRS which are potentially non-terminating.
Moreover, we also plan to analyze whether there are interesting
subclasses of PTRSs where $\AST$ or $\PAST$ is decidable.

To evaluate the power of our new techniques,
we used all 138 PTRSs from
the \emph{Termination Problem Data Base} \cite{TPDB}, i.e., the benchmarks from the annual
\emph{Termination Competition} \cite{termcomp}. They contain 138
typical probabilistic programs, including examples with complicated probabilistic
structure and probabilistic algorithms on lists and trees.
This set was mainly developed to evaluate techniques that can prove $\PASTASTColor$.
Therefore, most of the examples are indeed $\AST$, and we added 20 more examples that are not $\AST$ 
or not $\PAST$, including the PTRSs from this paper and examples
which express typical bugs in implementations.

{\footnotesize
\begin{center}
    \begin{tabular}{c @{\quad} c @{\quad} c @{\quad} c @{\quad} c @{\quad} c @{\quad} c
        @{\quad} c}
    \toprule
    Category & \Cref{theorem:embedding_loop_walks} & \Cref{theorem:embedding} & \Cref{theorem:embedding_two} & \Cref{theorem:embedding_three} & \Cref{theorem:embedding-pattern} & \aprove{} \\
    \midrule
    not-$\AST$ & 8~(2) & 17~(3) & 18~(4) & 16~(3) & 12~(5) & 24~(8) \\ 
    \midrule
    not-$\PAST$ & 8~(2) & 33~(7) & 34~(8) & 28~(6) & 30~(8) & 49~(14) \\ 
    \bottomrule
    \end{tabular}
\end{center}}

We performed our experiments on a computer with an Apple M4 CPU and 16 GB of RAM, 
and a timeout of 30 seconds was used for each example.
The table above shows the individual results for each of our novel theorems, and
``\aprove{}'' denotes the combination\footnote{So ``\aprove{}'' combines our five new
theorems for PTRSs. Without these theorems, \aprove{}
could only disprove $\PASTASTColor$
for examples that have no probabilities except 1.
}
of all techniques as implemented in our tool.
The numbers in brackets denote the results when just considering the 20 new examples.
Note that \aprove{} can \emph{prove}
$\AST$ for 70 of the 158 examples and \emph{prove} $\PAST$ for 31 of the 158 examples.
So at most $158 - 70 = 88$ examples may be non-$\AST$ and 
 \aprove{} can disprove $\AST$ for 24 of them. Similarly, at most $127$ examples may be
 non-$\PAST$ and \aprove{} disproves $\PAST$ for 49 of them.

The experimental results for \Cref{theorem:embedding_loop_walks}
confirm that one indeed needs
more elaborate techniques than just a
direct lifting of the loop detection technique to the
probabilistic setting. 
Our experiments  show that each of our five theorems has its own benefits. 
More precisely, for each  theorem, there exist examples that can only be
solved by this theorem but not by
any of the other four theorems.

Since ours is the first approach to disprove $\PASTASTColor$ of PTRSs automatically, we
could not compare  with other tools
for termination analysis of PTRSs. While there exist techniques
\cite{chatterjee2017StochasticInvariantsProbabilistic,takisaka2021RankingRepulsingSupermartingales}
and the tool \textsf{Amber} \cite{amber}
for disproving
$\PASTASTColor$ of imperative programs, an experimental comparison would be
problematic due to the fundamental differences between the considered languages.

For further details on our experiments 
and for instructions on how to run \textsf{AProVE} via its \emph{web interface}
or locally, we refer to:
\url{https://aprove-developers.github.io/DisprovingPTRSTermination/}

\subsubsection{Acknowledgments.} 
We thank Florian Frohn and Carsten Fuhs for a joint discussion
on initial ideas for this paper during CADE~'23 in Rome.

\begin{credits}
  \subsubsection{\discintname}
  The authors have no competing interests to declare.
\end{credits}

%% file: appendix.tex
\clearpage

\section{Appendix}\label{Appendix}
In this appendix, we give all proofs for our lemmas and theorems.

\EmbeddingLoopWalks*

\begin{myproof}
    We construct an infinite $\PP$-RST $\F{T}^{\infty}$ based on $\F{T}$ 
    where we always rewrite one of the occurrences of $t$ in a leaf according to the
    rules used to generate $\F{T}$.
    Since $\F{T}$ has an occurrence of $t$ in every leaf, the resulting tree $\F{T}^{\infty}$ has no leaves,
    and therefore, $\PP$ is neither $\AST$ nor $\PAST$.

    \begin{center}
        \begin{tikzpicture}[scale=0.9, transform shape]
        \tikzset{label distance=-2mm}
        \tikzstyle{empty}=[rectangle, thick,minimum size=1mm]
        \tikzstyle{adam}=[thick,draw=black!100,fill=white!100,minimum size=3mm,
            shape=rectangle split, rectangle split parts=2,rectangle split horizontal,font={\footnotesize}]

        \node[empty] at (-0.5, 0)  (TreeInf) {\large $\F{T}^{\infty}$:};
        \node[adam] at (0, -1)  (a) {$1$\nodepart{two}$\textcolor{purple}{t}$};

        \node[adam] at (3, 0)  (b) {$p_1$\nodepart{two}$C_1[\textcolor{purple}{t} \sigma_1]$};
        \node[adam] at (3, -1)  (c) {$p_2$\nodepart{two}$C_2[\textcolor{purple}{t} \sigma_2]$};
        \node[adam] at (3, -2)  (d) {$p_3$\nodepart{two}$C_3[\textcolor{purple}{t} \sigma_3]$};

        \node[empty] at (6, 0.25)  (b1) {\ldots};
        \node[empty] at (6, -0.25)  (b2) {\ldots};
        \node[empty] at (6, -0.75)  (c1) {\ldots};
        \node[empty] at (6, -1.25)  (c2) {\ldots};
        \node[empty] at (6, -1.75)  (d1) {\ldots};
        \node[empty] at (6, -2.25)  (d2) {\ldots};

        \draw (a) edge[dashed, -{Stealth}] (b);
        \draw (a) edge[dashed, -{Stealth}] (c);
        \draw (a) edge[dashed, -{Stealth}] (d);

        \draw (b) edge[dashed, -{Stealth}] (b1);
        \draw (b) edge[dashed, -{Stealth}] (b2);
        \draw (c) edge[dashed, -{Stealth}] (c1);
        \draw (c) edge[dashed, -{Stealth}] (c2);
        \draw (d) edge[dashed, -{Stealth}] (d1);
        \draw (d) edge[dashed, -{Stealth}] (d2);
    \end{tikzpicture}
    \end{center}

    We only have to show that if we rewrite a term $s = C[t \sigma]$ for some context $C$ and some substitution $\sigma$
    using the same rules as in $\F{T}$, then we obtain a tree $\F{T}_{s}$ where all leaves contain an occurrence of $t$ again.
    Let $\tau$ be the position of the hole in $C$.
    If we rewrite at position $\pi$ in $\F{T}$, then we rewrite at position $\tau.\pi$ in the tree $\F{T}_{s}$.
    For every leaf $v \in \ctleaf(\F{T})$ there is a corresponding leaf $v' \in
    \ctleaf(\F{T}_{s})$
    with $t_{v'} = C[t_{v} \sigma]$.
    Since there exists an occurrence of $t$ within $t_{v}$,
    i.e., $t_{v} = C'[t \sigma']$, there is an occurrence of $t$ within $t_{v'} = C[t_{v} \sigma] = C[C'[t \sigma'] \sigma]$ as well.
\end{myproof}

Before we can prove \Cref{theorem:lower_bounds_by_embeddings},
we show two auxiliary lemmas. They are concerned with embeddings from
an RST $\F{T}_1$ to an RST $\F{T}_2$.
Here,    let $\pre(v)$ be the set of all (not necessarily direct) 
    predecessors of a node $v \in V(\F{T}_2)$ 
    including the node $v$ itself.
  Similarly,  let $\post(v)$ be the set of all (not necessarily direct) 
successors of a node $v \in V(\F{T}_2)$ 
including the node $v$ itself.
By $\post_{\ctleaf}(v)$ we denote the set of all leaves in $\post(v)$.

\begin{lemma}\label{lemma:disjoint-leaf-images}
    Let ${(A_1, \to_1)}$, ${(A_2, \to_2)}$ be two PARSs, and let $\F{T}_i$ be a $\to_i$-RST for $i \in \{1,2\}$
    such that there exists an embedding from $\F{T}_1$ to $\F{T}_2$.
    Let $u_1, u_2 \in \ctleaf(\F{T}_1)$ with $u_1 \neq u_2$.
Then, the sets $\post_{\ctleaf}(\mathtt{e}(u_1))$ and
$\post_{\ctleaf}(\mathtt{e}(u_2))$ are disjoint.
\end{lemma}

\begin{myproof}
    Assume for a contradiction that there is a leaf $w \in \ctleaf(\F{T}_2)$ with
    $w \in \post_{\ctleaf}(\mathtt{e}(u_1)) \cap \post_{\ctleaf}(\mathtt{e}(u_2))$.
    Then both $\mathtt{e}(u_1)$ and $\mathtt{e}(u_2)$ are predecessors of $w$, and therefore,
    one of them is above the other, i.e., there is a path from $\mathtt{e}(u_1)$ to $\mathtt{e}(u_2)$
or from $\mathtt{e}(u_2)$ to $\mathtt{e}(u_1)$ in $\F{T}_2$.

    W.l.o.g., let $\mathtt{e}(u_1)$ be above $\mathtt{e}(u_2)$.
    Let $v_0 \to v_1 \to \cdots \to v_n = u_2$ be the path in $\F{T}_1$ from the root to $u_2$.
    By path preservation of the embedding $\mathtt{e}$, for every $i < n$ there is a path from $\mathtt{e}(v_i)$ to
    $\mathtt{e}(v_{i+1})$ in $\F{T}_2$, and each such path does not contain any other
    nodes from $\mathtt{e}(V(\F{T}_1))$ by
    \Cref{def:embedding}.

    Since $\mathtt{e}(u_1)$ is a strict predecessor of $\mathtt{e}(u_2)$,
    there is a smallest index $i$ such that $\mathtt{e}(u_1)$ is a predecessor of
    $\mathtt{e}(v_{i+1})$.
    By minimality, $\mathtt{e}(u_1)$ is not a predecessor of $\mathtt{e}(v_i)$.
    Hence, $\mathtt{e}(u_1)$ is an inner node on the path from
    $\mathtt{e}(v_i)$ to $\mathtt{e}(v_{i+1})$.
    As $u_1 \neq v_i$ and $u_1 \neq v_{i+1}$, injectivity yields
    $\mathtt{e}(u_1) \neq \mathtt{e}(v_i)$ and $\mathtt{e}(u_1) \neq \mathtt{e}(v_{i+1})$.
    This contradicts the direct-successor condition in \Cref{def:embedding}.
\end{myproof}

\begin{lemma}\label{lemma:leaf-predecessor-image}
    Let ${(A_1, \to_1)}$, ${(A_2, \to_2)}$ be two PARSs, and let $\F{T}_i$ be a $\to_i$-RST for $i \in \{1,2\}$
    such that there exists an embedding from $\F{T}_1$ to $\F{T}_2$.
   Then for every $w \in \ctleaf(\F{T}_2)$, there is a predecessor $v \in \pre(w)$ and a
    leaf $u \in \ctleaf(\F{T}_1)$ such that $v = \mathtt{e}(u)$.
\end{lemma}

\begin{myproof}
    We start with some basic observations.
    \begin{enumerate}[label=\textcolor{purple}{\Roman*.}, leftmargin=*, widest=III.]
        \item Since we have $p_v = p_{\mathtt{e}(v)}$ for all $v \in V(\F{T}_1)$ 
        by definition of an embedding (\Cref{def:embedding}) 
        and $p_{\ctroot} = 1$ for the root node $\ctroot$ of $\F{T}_1$
        by definition of a $\to$-RST,
        we must map the root $\ctroot$ of $\F{T}_1$ to some node $v \in V(\F{T}_2)$
        with $p_{v} = p_{\mathtt{e}(\ctroot)} = p_{\ctroot} = 1$.

        \item The nodes $v_1, \ldots, v_n$ of  $\F{T}_2$ with probability $1$,
        i.e., $v \in V(\F{T}_2)$ with $p_{v} = 1$, form a connected path.
        Let $v'$ be the node with the greatest depth within this path.
        Every leaf $w \in \ctleaf(\F{T}_2)$ is a successor of $v'$, i.e., $w \in \post(v')$. 
        Thus, together with \textcolor{purple}{I.} we obtain that every leaf $w \in \ctleaf(\F{T}_2)$ 
        has a predecessor $v \in \pre(w)$ such that $v = \mathtt{e}(u)$ for some $u \in V(\F{T}_1)$.
    \end{enumerate}

    It remains to show that for every leaf $w \in \ctleaf(\F{T}_2)$, 
    there is a predecessor $v \in \pre(w)$ and a
    \emph{leaf $u \in \ctleaf(\F{T}_1)$} such that $v = \mathtt{e}(u)$.
    Thus, let $w \in \ctleaf(\F{T}_2)$.
    By Observation \textcolor{purple}{II.}, there is some predecessor of $w$ in the image of
    $\mathtt{e}$. We choose such a predecessor $v = \mathtt{e}(u)$ with maximal depth.
    Assume that $u$ is not a leaf of $\F{T}_1$.
    Let $u_1,\ldots,u_k$ be the direct successors of $u$ in $\F{T}_1$.
    By path preservation, all $\mathtt{e}(u_j)$ are successors of $v$.
    Moreover,
    \[
    p_v = p_{\mathtt{e}(u)} = p_u = \sum_{j=1}^{k} p_{u_j} = \sum_{j=1}^{k} p_{\mathtt{e}(u_j)}.
    \]
    Thus, the whole probability mass below $v$ is exhausted by the subtrees of
$\F{T}_2$ with the roots
    $\mathtt{e}(u_1),\ldots,\mathtt{e}(u_k)$.
    Since $w$ is a leaf below $v$, it must belong to one of these subtrees, i.e.,
    $\mathtt{e}(u_j) \in \pre(w)$ for some $j$.
    But this contradicts the maximality of the depth of $v = \mathtt{e}(u)$.
    Hence, $u$ is a leaf.
\end{myproof}

\LowerBounds*

\begin{myproof}
   First, we show $|\F{T}_2| \leq |\F{T}_1|$.
    If $v \in \ctleaf(\F{T}_2)$,
    then by \Cref{lemma:leaf-predecessor-image} there has to exist a $w \in \pre(v)$ 
    that is the image of a leaf $u \in \ctleaf(\F{T}_1)$ in $\F{T}_1$ (i.e.,
    $\mathtt{e}(u) = w$).
    Note that by the definition of embeddings we have $p_{w} = p_{\mathtt{e}(u)} = p_{u}$.
    Moreover, for distinct leaves $u_1, u_2 \in \ctleaf(\F{T}_1)$,
    the sets $\post_{\ctleaf}(\mathtt{e}(u_1))$ and $\post_{\ctleaf}(\mathtt{e}(u_2))$ are disjoint
    by \Cref{lemma:disjoint-leaf-images}.
    Furthermore, we have $\sum_{v \in \post_{\ctleaf}(w)} p_{v} \leq p_{w}$ for every node $w$.
    Thus, we obtain
    \[\begin{array}{r@{\;\;}c@{\;\;}l@{\;\;}c@{\;\;}l@{\;\;}c@{\;\;}l}
        |\F{T}_2| &=& \sum_{v \in \ctleaf(\F{T}_2)} p_{v} 
        &=& \sum_{u \in \ctleaf(\F{T}_1)} \sum_{v \in \post_{\ctleaf}(\mathtt{e}(u))} p_{v} \\
        &\leq&
        \sum_{u \in \ctleaf(\F{T}_1)} p_{\mathtt{e}(u)} &=&
        \sum_{u \in \ctleaf(\F{T}_1)} p_{u} &=& |\F{T}_1|.
    \end{array}\]

    Next, we consider the expected derivation length. If $|\F{T}_2| < 1$, then we have
    $\edl(\F{T}_2) = \infty$ and thus, the claim is trivial. Hence, let $|\F{T}_2| = 1$.
    Due to $|\F{T}_2| \leq |\F{T}_1|$,
    this also implies $|\F{T}_1| = 1$.
    Note that a path from the root to a leaf $v$ in $\F{T}_2$
    is at least as long as the path from the root to the corresponding leaf $u$ in
    $\F{T}_1$
    (where $\mathtt{e}(u) = w$ for a $w \in \pre(v)$),
    since $\mathtt{e}$ is injective and ``path preserving''.
    Therefore, we have $\ctdepth(u) \leq \ctdepth(v)$ for all $v \in \post_{\ctleaf}(\mathtt{e}(u))$,
    and thus      
    \[\begin{array}{rcl}
        \edl(\F{T}_2) 
        &=& \sum_{u \in \ctleaf(\F{T}_1)} \sum_{v \in \post_{\ctleaf}(\mathtt{e}(u))} \ctdepth(v) \cdot p_{v} \\
        &\geq&
      \sum_{u \in \ctleaf(\F{T}_1)} \sum_{v \in
          \post_{\ctleaf}(\mathtt{e}(u))} \ctdepth(u) \cdot p_v\\
              &=&     \sum_{u \in \ctleaf(\F{T}_1)} \ctdepth(u) \cdot \sum_{v \in
        \post_{\ctleaf}(\mathtt{e}(u))} p_v\\
        &=& \sum_{u \in \ctleaf(\F{T}_1)} \ctdepth(u) \cdot  p_{\mathtt{e}(u)} \\ &=&
        \sum_{u \in \ctleaf(\F{T}_1)} \ctdepth(u) \cdot p_{u} \\
        &=& \edl(\F{T}_1)
        \end{array}
        \]
                To conclude  $\sum_{v \in
                    \post_{\ctleaf}(\mathtt{e}(u))} p_v = p_{\mathtt{e}(u)}$ in the fourth step, note that
                $|\F{T}_2| = 1$ implies that we have $p_w = \sum_{v \in \post_{\ctleaf}(w)} p_v$ for every
        node $w$ of  $\F{T}_2$.
\end{myproof}

\EmbeddingRW*

\begin{myproof}
    If we have $|\F{T}| = \sum_{v \in \ctleaf(\F{T})} p_v < 1$, 
    then $\F{T}$ is itself a witness that $\PP$ is not $\AST$ and not $\PAST$.
    Therefore, we now consider the case $|\F{T}| = 1$.
    
    We define the random walk $\mu$ by 
    $\mu(x) = \sum_{v \in \ctleaf(\F{T}), x = \maxNOO(t,t_v)-1} p_v$ for all $x \in \IZ$.
    Note that
    \[\begin{array}{rcl}
    \IE(\mu) &=& \sum_{v \in \ctleaf(\F{T})} p_v \cdot (\maxNOO(t,t_v) -1)\\
    &=& (\sum_{v \in \ctleaf(\F{T})} p_v \cdot \maxNOO(t,t_v)) -
    (\sum_{v \in \ctleaf(\F{T})} p_v)\\
    &=& (\sum_{v \in \ctleaf(\F{T})} p_v \cdot \maxNOO(t,t_v)) - 1 \qquad \text{\textcolor{blue}{(since $\sum_{v
        \in \ctleaf(\F{T})} p_v = 1$)}}
    \end{array}\]
    Therefore, if $\sum_{v \in \ctleaf(\F{T})} p_v \cdot \maxNOO(t,t_v) > 1$, then we have
    $\IE(\mu) > 0$ and by \Cref{theorem:termination}, $\mu$ is not $\AST$.
    Similarly, if $\sum_{v \in \ctleaf(\F{T})} p_v \cdot \maxNOO(t,t_v) \geq 1$, then we have
    $\IE(\mu) \geq 0$ and by \Cref{theorem:termination}, $\mu$ is not $\PAST$.

    We can represent $\mu$ as a PARS $(\IZ, \hookrule_{\mu})$ with 
    $y \hookrule_{\mu} \{(p:y+\maxNOO(t,t_v)-1) \mid v \in \ctleaf(\F{T})\}$ for every $y > 0$.
    Let $\F{T}_1$ be the infinite $\hookrule_{\mu}$-RST that starts at $1$.
    Note that $|\F{T}_1| < 1$ if $\mu$ is not $\AST$, 
    and $\edl(\F{T}_1) = \infty$ if $\mu$ is not $\PAST$.
    The only difference between $\mu$ and the PARS $(\IZ, \hookrule_{\mu})$ 
    is that $\hookrule_{\mu}$ is defined via multi-distributions which
    may have multiple different pairs $(p_1:x), \ldots, (p_k:x)$
    for the same number $x$, while $\mu$ maps every value $x$ to
    a single probability ($\mu(x) = \sum_{i = 1}^k p_i$).
    However, since for every number $x$ there is only a single rewrite rule with $x$ as
    its left-hand side, 
    the termination probability and expected derivation length are equal for $\mu$ and $(\IZ, \hookrule_{\mu})$.

    We construct an infinite $\PP$-RST $\F{T}^{\infty}$ based on $\F{T}$ 
    where we always rewrite an innermost occurrence of $t$ in a leaf according to the
    rules used to generate $\F{T}$.
    If we have $\maxNOO(t,t_v) = 0$ in a leaf $v$, then it remains a leaf in $\F{T}^{\infty}$.
    We have to prove that $\F{T}^{\infty}$ behaves like the rewrite sequence tree of our random walk $\mu$. 

    We first show that if we have a term $s \in \TT$ with $\maxNOO(t,s) = k \in \IN_{> 0}$,
    then instead of the tree $\F{T}$ whose root is labeled with $t$, we can construct a
    tree $\F{T}_s$ whose root is labeled with $s$, where  $\F{T}_s$ is like $\F{T}$ when
    rewriting an innermost occurrence of  $t$ in $s$ instead, according to the rules used in $\F{T}$.
    Thus, we can define a bijective embedding $\mathtt{e}_s: V(\F{T}) \to V(\F{T}_s)$
    and obtain
    $\maxNOO(t,t_{\mathtt{e}_s(v)}) \geq  \maxNOO(t,t_{v}) + k - 1$ for all leaves $v \in \ctleaf(\F{T})$.
    (While $k$ was the original number of  occurrences of $t$ in $s$, this number is now
    modified according to the random walk $\mu$.)
    To prove this inequation, let
    $\{\pi_1, \ldots, \pi_k\} \in \NOO(t,s)$ be a witness for the $k$ pairwise
    non-overlapping occurrences of $t$ in $s$.
    The term $s$ has the form $s = C[t\delta]$ 
    for some context $C$ with a hole at position $\pi_1$ and substitution $\delta$.
    Let $\pi_1$ be the position of an innermost occurrence of $t$, i.e., 
    there is no $\pi_j$ with $1 \leq j \leq k$ strictly below $\pi_1$.
    To construct $\F{T}_s$, we
    rewrite $C[t\delta]$ according to $\F{T}$ at position $\pi_1$.
    This is possible, since if $t$ can be rewritten via the rules used in $\F{T}$ then so
    does every instantiation of $t$. 
    Let $v \in \ctleaf(\F{T})$ and $\mathtt{e}_s(v) \in \ctleaf(\F{T}_s)$ be the corresponding leaf in $\F{T}_s$
    which we get by following the same path in $\F{T}_s$ as in $\F{T}$.
    We have $t_{\mathtt{e}_s(v)} = C[t_v\delta]$.
    Since $t$ is linear and $\{\pi_1, \ldots, \pi_k\}$ are pairwise non-overlapping w.r.t.\ $t$,
    we still have $\{\pi_2, \ldots, \pi_k\} \in \NOO(t,C[t_v\delta])$.
    Moreover, all occurrences of $t$ in $t_v$ also exist in $t_v\delta$.
    Finally, the occurrences of $t$ in $t_v\delta$ are non-overlapping with
    $\{\pi_2, \ldots, \pi_k\}$ because they are below $\pi_1$.
    Thus, we get $\maxNOO(t,t_{\mathtt{e}_s(v)}) = \maxNOO(t,C[t_v\delta]) = \maxNOO(t,t_v\delta) + k - 1 \geq \maxNOO(t,t_v) + k - 1$.

    Now we define the embedding $\mathtt{e} : V(\F{T}_1) \to V(\F{T}^{\infty})$
    from the $\hookrule_{\mu}$-RST 
    by induction on the depth of the node in $\F{T}_1$. Moreover, for every node
    $w \in V(\F{T}_1)$ that is labeled with the number $a_w$ in $\F{T}_1$, we prove  $a_w  \leq \maxNOO(t,t_{\mathtt{e}(w)})$.
     
    We start by mapping the root $w$ of $\F{T}_1$ to the root of $\F{T}^{\infty}$. Here,
    we have $a_w = 1$ and $\maxNOO(t,t_{\mathtt{e}(w)}) = \maxNOO(t,t) = 1$.

    Now assume that we have already defined the mapping for a node $w \in V(\F{T}_1)$ and
    we have  $a_w  \leq \maxNOO(t,t_{\mathtt{e}(w)})$. Every child $u$ of  $w$ in $V(\F{T}_1)$
    is labeled by a number of the form 
    $a_u = a_w+x$ with $x \in \IZ$ and $\mu(x) > 0$. Thus, by the definition of $\mu$, there is a
    $v \in \ctleaf(\F{T})$ with $x = \maxNOO(t,t_v) -1$. Consider the tree
    $\F{T}_{t_{\mathtt{e}(w)}}$. As  $a_w  \leq \maxNOO(t,t_{\mathtt{e}(w)})$, we know
    that if $w$ has a child in $\F{T}_1$, then $1 \leq a_w \leq \maxNOO(t,t_{\mathtt{e}(w)})$.
    Thus, by the inequation proved above,
    there is a bijective embedding $\mathtt{e}_{t_{\mathtt{e}(w)}}$ of $\F{T}$ into
    $\F{T}_{t_{\mathtt{e}(w)}}$ where
    $\maxNOO(t,t_{\mathtt{e}_{t_{\mathtt{e}(w)}}(v)}) \geq  \maxNOO(t,t_v) + k - 1$ for $k =
    \maxNOO(t,t_{\mathtt{e}(w)})$. As $a_w \leq k$, this implies
    $\maxNOO(t,t_{\mathtt{e}_{t_{\mathtt{e}(w)}}(v)})\geq  \maxNOO(t,t_v) + a_w - 1$.
    In $\F{T}^{\infty}$, the node $\mathtt{e}(w)$ eventually reaches a node
    corresponding to $\mathtt{e}_{t_{\mathtt{e}(w)}}(v)$. 
    Thus, $\mathtt{e}(u)$ is defined to be this node.
    Then we have $a_u = a_w+x = a_w + \maxNOO(t,t_v) -1 \leq
    \maxNOO(t,t_{\mathtt{e}_{t_{\mathtt{e}(w)}}(v)}) =    \maxNOO(t,t_{\mathtt{e}(u)})$.
    Injectivity, path preservation, and equal probabilities for the embedding $\mathtt{e}$ follow by construction.

    By \Cref{theorem:lower_bounds_by_embeddings}, 
    we get $|\F{T}^{\infty}| \leq |\F{T}_1|$ and $\edl(\F{T}^{\infty}) \geq \edl(\F{T}_1)$.
    So if $|\F{T}_1| < 1$ holds, then $\PP$ is not $\AST$, and if $\edl(\F{T}_1) = \infty$ holds, then $\PP$ is not $\PAST$.
\end{myproof}

\EmbeddingRWTwo*

\begin{myproof}
    The proof is analogous to the one of \Cref{theorem:embedding},
    but now we construct the infinite $\PP$-RST $\F{T}^{\infty}$ based on $\F{T}$ 
    by always rewriting an outermost occurrence of $t$.

    Again,  if  $s \in \TT$ with $\maxNOO(t,s) = k \in \IN_{> 0}$,
    then instead of the tree $\F{T}$ whose root is labeled with $t$, we show that we
    can construct a
    tree $\F{T}_s$ whose root is labeled with $s$, where  $\F{T}_s$ is like $\F{T}$ when
    rewriting an outermost occurrence of  $t$ in $s$ instead, according to the rules used in $\F{T}$.
    As in the proof of \Cref{theorem:embedding},
    we can define a bijective embedding $\mathtt{e}_s: V(\F{T}) \to V(\F{T}_s)$
    and obtain
    $\maxNOO(t,t_{\mathtt{e}_s(v)}) \geq  \maxNOO(t,t_{v}) + k - 1$ for all leaves $v \in \ctleaf(\F{T})$.
    To prove this inequation, 
    let $\{\pi_1, \ldots, \pi_k\} \in \NOO(t,s)$ be a witness for the $k$ pairwise
    non-overlapping occurrences in $s$.
    The term $s$ has the form $s = C[t\delta]$ 
    for some context $C$ with a hole at position $\pi_1$ and substitution $\delta$.
    Let $\pi_1$ be the position of an outermost occurrence of $t$, i.e.,
    there is no $\pi_j$ with $1 \leq j \leq k$ strictly above $\pi_1$.
    We rewrite $C[t\delta]$ according to $\F{T}$ at position $\pi_1$.
    Since $\F{T}$ is non-variable-decreasing, 
    for every position $\tau \in \pos_{\VSet}(t)$
    and every leaf $v \in \ctleaf(\F{T})$,
    there exists a (unique) position $\tau' \in \pos_{\VSet}(t_v)$ 
    with $t|_{\tau} = {t_v}|_{\tau'}$. 

    As in the proof of \Cref{theorem:embedding}, we get $t_{\mathtt{e}_s(v)} = C[t_v\delta]$.
    All occurrences of $t$ at positions
    that are orthogonal to $\pi_1$ are still present in $t_{\mathtt{e}_s(v)}$.
    All occurrences at a position below $\pi_1$, i.e., $\pi_j = \pi_1.\kappa$,
    can be written as $\pi_j = \pi_1.\tau.\beta$ for some position $\beta \in \IN^*$ and
    $\tau \in \pos_{\VSet}(t)$.
    Moreover, since $t \subtermocAlt_{\pi_j} C[t\delta]$ we get $t \subtermocAlt_{\beta} \delta(x)$
    for the variable $x = t|_{\tau}$.
    By the previous paragraph, we can find a (unique) position $\tau' \in \pos_{\VSet}(t_v)$ 
    with $t|_{\tau} = {t_v}|_{\tau'}$, and therefore, all such occurrences are still
    present in
    $t_{\mathtt{e}_s(v)} = C[t_v\delta]$, because
    we still have $t \subtermocAlt_{\pi_1.\tau'.\beta} C[t_v\delta] \Longleftrightarrow t \subtermocAlt_{\beta} \delta(x)$
    for the variable $x = t_v|_{\tau'} = t|_{\tau}$.
    Note that there are no occurrences above $\pi_1$ as it is an outermost occurrence.
    Thus, we obtain $\{\pi_2, \ldots, \pi_k\} \in \NOO(t,C[t_v\delta])$.
    Moreover, all occurrences of $t$ in $t_v$ also exist in $t_v \delta$. Finally,
    the occurrences of $t$ in $t_v$ are non-overlapping with
    $\{\pi_2, \ldots, \pi_k\}$. 
    Hence, we get $\maxNOO(t,t_{\mathtt{e}_s(v)}) = \maxNOO(t,C[t_v\delta]) 
    = \maxNOO(t,t_v\delta) + k - 1 \geq \maxNOO(t,t_v) + k - 1$ as desired.
    The rest of the proof (i.e., the definition of
    the embedding 
    from the $\hookrule_{\mu}$-RST to $\F{T}^{\infty}$) is as in the proof
    of \Cref{theorem:embedding}.
\end{myproof}

\EmbeddingRWThree*

\begin{myproof}
    Compared to the proof of \Cref{theorem:embedding}, 
    we now define the random walk $\mu$ via $\maxParaNOO(t,t_v)$ instead of $\maxNOO(t,t_v)$.
    The remaining proof only differs in the proof
    of the inequation $\maxParaNOO(t,t_{\mathtt{e}_s(v)}) \geq
    \maxParaNOO(t,t_v) + k - 1$, where  $\maxParaNOO(t,s) = k \in \IN_{> 0}$.
    
    Let $\{\pi_1, \ldots, \pi_k\} \in \NOO(t,s)$ be a witness for the 
    $k$ orthogonal occurrences of $t$ in $s$.
    The term $s$ has the form $s = C[t\delta]$ 
    for some context $C$ with a hole at position $\pi_1$ and substitution $\delta$.
    We rewrite $C[t\delta]$ according to $\F{T}$ at position $\pi_1$, and get $t_{\mathtt{e}_s(v)} = C[t_v\delta]$.

    All the occurrences of $t$ at the positions
    $\pi_2, \ldots, \pi_k$ are still present,
    because positions orthogonal to $\pi_1$ remain in $C[t_v\delta]$.
    Moreover, all orthogonal occurrences of $t$ in $t_v\delta$ are orthogonal with all occurrences 
    at a position in $\{\pi_2, \ldots, \pi_k\}$ because they are below $\pi_1$.
    Hence, we get $\maxParaNOO(t,t_{\mathtt{e}_s(v)}) = \maxParaNOO(t,C[t_v\delta]) 
    = \maxParaNOO(t,t_v\delta) + k - 1 \geq \maxParaNOO(t,t_v) + k - 1$ as desired.
\end{myproof}

\SuffAndNecForInequal*

\begin{myproof}
  We first show the ``if'' direction, i.e., we prove that the condition of \Cref{lemma:suff_and_nec_for_inequality} is
  sufficient. To this end, we first show that $x, x \sigma, x \sigma^2, \ldots$ are
  all pairwise different.

  Since $x$ is on a cycle of  $G_{\sigma, t}$, there exists a minimal $k \in \IN_{> 0}$ such that $x
  \in \VSet(x\sigma^k)$. 
  
  If $k = 1$, then we have  $x  \in \VSet(x\sigma)$, i.e., $x \sigma = C[x]$ for a non-empty
  context $C \neq \Box$. Thus, all $x \sigma^m = C^m[x]$ are pairwise different for $m
  \geq 0$.

  Otherwise, if $k > 1$, first note that $x, x \sigma, \ldots, x \sigma^{k-1}$ are
  pairwise different. The reason is that otherwise, there would be $0 \leq i < j \leq k-1$
  with $x \sigma^i = x \sigma^j$. But this would imply $x \sigma^{i+k-j} = x \sigma^{j+
    k-j} = x\sigma^k$, which is a contradiction because
$x \in \VSet(x\sigma^k)$, but
  $x \notin \VSet(x \sigma^{i+k-j})$ due to the minimality of $k$.

  Next, note that $x\sigma^k, x \sigma^{k+1}, \ldots, x \sigma^{2\cdot k-1}$ are also
  pairwise different. The reason is that $x\sigma^k = C[x]_\pi$ for a non-empty context $C \neq
  \Box$. Hence, the subterms of $x\sigma^k, x \sigma^{k+1}, \ldots, x \sigma^{2\cdot k-1}$ 
at position $\pi$ are  $x, x
  \sigma, \ldots, x \sigma^{k-1}$ which are pairwise different by the observation
  above. Moreover,  all terms $x\sigma^k, x \sigma^{k+1}, \ldots, x \sigma^{2\cdot k-1}$
  are pairwise different from the terms $x, x \sigma, \ldots, x \sigma^{k-1}$ due to the
  additional non-empty contexts.
  By repeating this reasoning, we can infer that all terms $x, x \sigma, x \sigma^2,
  \ldots$ are pairwise different.

  Now we need to show that all terms $t, t \sigma, t \sigma^2, \ldots$ are pairwise
  different. By the definition of $G_{\sigma, t}$, $t$ contains a variable $y$ at a
  position $\tau$ where $y$ has a path
  to $x$ in $G_{\sigma, t}$. Let $j$ be the length of the shortest path from $y$ to $x$, i.e., $j\geq 0$ is
  the minimal number such that $x \in \VSet(y\sigma^j)$. Similar to the argumentation above,
  this implies that $y, y\sigma, \ldots, y\sigma^j$ are pairwise different. By considering
  only the subterm of $t$ at position $\tau$, this also means that $t, t\sigma, \ldots,
  t\sigma^j$ are pairwise different. As $t\sigma^j$ contains the variable $x$ at some
  position, by considering the subterm at this position and the argumentation above, we
  obtain that all terms $t, t\sigma, \ldots,
  t\sigma^j, t\sigma^{j+1}, \ldots$ are pairwise different.

  \medskip

  Now we show the ``only if'' direction, i.e., we prove that the condition of \Cref{lemma:suff_and_nec_for_inequality} is
  necessary. We assume the contrary, i.e., assume that for all variables $x$ in cycles of
  $G_{\sigma, t}$ we have $x\sigma \in \VSet$. This means that for every such variable $x$
  there exists a $k_x > 0$ such that $x \sigma^{k_x} = x$.

  Moreover, $x\sigma \in \VSet$
implies that
  every variable in a cycle of  $G_{\sigma, t}$ only has one outgoing edge.
  Therefore,
  $t \sigma^{|\dom(\sigma)|}$ only contains variables on cycles of $G_{\sigma, t}$. Let $k
  = k_{x_1} \cdot \ldots \cdot k_{x_n}$ where $x_1, \ldots, x_n$ are all variables on
  cycles of  $G_{\sigma, t}$. Thus, for all $m \in \IN$ we have
  $t \sigma^{|\dom(\sigma)|} = t \sigma^{|\dom(\sigma)| + m \cdot k}$, which shows that
  $\langle t,\sigma \rangle$ is not a pattern term.
  \end{myproof}

\EmbeddingRWPattern*

\begin{myproof}
    Compared to the proof of \Cref{theorem:embedding_three}, 
    we define the random walk $\mu$ via $\maxParaNOM(t, \sigma, t_v)$ instead of $\maxParaNOO(t,t_v)$.
    So we now have to prove the inequation $\maxParaNOM(t, \sigma, t_{\mathtt{e}_s(v)})
    \geq \maxParaNOM(t, \sigma, t_v) + k - 1$, where $\maxParaNOM(t, \sigma, s) = k > 0$.

    Let $\{\pi_1, \ldots, \pi_h\} \in \NOO(t,s)$ be a witness 
    for $h$ orthogonal occurrences in $s$
    where the sum of multiplicities is $k$.
    The term $s$ has the form $s = C[t \sigma^{m_{\pi_1}} \delta]$ 
    for some context $C$ with a hole at position $\pi_1$ and some substitution $\delta$,
    where $m_{\pi_1} > 0$.
    Let $\F{T}_{s}$ be the tree resulting from starting with $s=
    C[t \sigma^{m_{\pi_1}} \delta]$ at the root and by rewriting
    $t \sigma^{m_{\pi_1}} \delta$ at position $\pi_1$
    according to the rules used in the tree $\F{T}$.
    For every $v \in \ctleaf(\F{T})$ let
    $\mathtt{e}_s(v) \in \ctleaf(\F{T}_s)$ 
    be the corresponding leaf in $\F{T}_s$
    which we get by following the same path in $\F{T}_s$ as the path to the leaf
    $v$ in $\F{T}$.
    We have $t_{\mathtt{e}_s(v)} = C[t_{v} \sigma^{m_{\pi_1} - 1} \delta] = C[t_{v'} \delta]$
    for the corresponding leaf $v'$ in the tree $\F{T}_{m_{\pi_1}}$ that starts with $t \sigma^{m_{\pi_1}}$
    and uses the same rules as in $\F{T}$ at the same positions.
    
    All the occurrences of $\langle t, \sigma \rangle$ at the positions
    $\pi_2, \ldots, \pi_h$ with multiplicities $m_{\pi_2}, \ldots, m_{\pi_h}$ are still present,
    because they are all pairwise orthogonal.
    Let $\{\chi_1, \ldots, \chi_p\}$ be a witness 
    for $p$ orthogonal occurrences in $t_v$
    where the sum of multiplicities is $\maxParaNOM(t,\sigma,t_v)$.
    By the requirements on pattern trees, we have $p \geq 1$.
    Due to commutation, for every occurrence $\langle t, \sigma \rangle \subtermocAlt^{m_{\chi_j}}_{\chi_j} t_v$ 
    for a leaf $v \in \ctleaf(\F{T})$ and every $\kappa_{\chi_j, v}$ such that 
    $t \sigma^{m_{\chi_j}} \kappa_{\chi_j, v} = t_v|_{\chi_j}$
    we have a (unique) occurrence
    $\langle t, \sigma \rangle \subtermocAlt^{m_{\chi_j} + m_{\pi_1}-1}_{\chi_j} t_{v'}$
    in the corresponding leaf $v' \in \F{T}_{m_{\pi_1}}$.
    To see this, note that 
    \[t_{v'}|_{\chi_j} = t_v|_{\chi_j} \sigma^{m_{\pi_1}-1} 
    = t \sigma^{m_{\chi_j}} \kappa_{\chi_j, v} \sigma^{m_{\pi_1}-1} 
    = t \sigma^{m_{\chi_j} + m_{\pi_1}-1} \kappa_{\chi_j, v}.\]
    \pagebreak[3]
    
    We get 
     \begin{align*}
         & \maxParaNOM(t,\sigma,t_{v'}) \\
        = & \max\{m_S \mid S \in \NOO(t, t_{v'}), \forall \pi_1, \pi_2 \in S 
        \text{ with }\pi_1 \neq \pi_2: \pi_1 \bot \pi_2\}\\
        = & \max\{\sum_{\chi \in S} m_{\chi} \mid S \in \NOO(t, t_{v'}), \forall \pi_1, \pi_2 \in S 
        \text{ with }\pi_1 \neq \pi_2: \pi_1 \bot \pi_2\}\\
         \textcolor{blue}{\big\Downarrow} & \quad \textcolor{blue}{\bigl( \text{since } \{\chi_1, \ldots, \chi_p\} \text{ is a set of orthogonal occurrences} \bigr)}\\
         \geq & \sum_{\chi \in \{\chi_1, \ldots, \chi_p\}} m_{\chi}\\
         \textcolor{blue}{\big\Downarrow} & \quad \textcolor{blue}{\bigl( \text{by the commutation property of } \F{T} \text{ as described above}\bigr)}\\
         = & \sum_{\chi \in \{\chi_1, \ldots, \chi_p\}} \bigl(m_{\chi} + m_{\pi_1} -
         1\bigr)\\
    = & \bigl(\sum_{\chi \in \{\chi_1, \ldots, \chi_p\}} m_{\chi} \bigr) + p \cdot (m_{\pi_1} -
    1)\\
      \textcolor{blue}{\big\Downarrow} & \quad \textcolor{blue}{\bigl( \text{since $p \geq
          1$ and $m_{\pi_1} \geq 1$}\bigr)}\\
        \geq & \bigl(\sum_{\chi \in \{\chi_1, \ldots, \chi_p\}} m_{\chi}\bigr) + m_{\pi_1} - 1\\
         \textcolor{blue}{\big\Downarrow} & \quad \textcolor{blue}{\bigl( \text{since } \{\chi_1, \ldots, \chi_p\} \text{ is a witness of the maximum} \bigr)}\\
                 = & \max\{m_S \mid
S \in \NOO(t, t_v),
\forall \pi_1, \pi_2 \in S \text{
  with }\pi_1 \neq \pi_2: \pi_1 \bot \pi_2\} + m_{\pi_1} - 1\\
         = & \maxParaNOM(t,\sigma,t_v) + m_{\pi_1} - 1. 
     \end{align*}
    Overall, because of $t_{\mathtt{e}_s(v)} = C[t_{v'} \delta]$ we get
    \begin{align*}
        & \maxParaNOM(t,\sigma,t_{\mathtt{e}_s(v)}) \\
        = & \maxParaNOM(t,\sigma,C[t_{v'} \delta])  \\
        \textcolor{blue}{\big\Downarrow} & \quad \textcolor{blue}{\bigl( \text{removing } \delta \text{ can only decrease the number of occurrences} \bigr)}\\
        \geq & \maxParaNOM(t,\sigma,C[t_{v'}])  \\
        \textcolor{blue}{\big\Downarrow} & \quad \textcolor{blue}{\bigl( \text{since the occurrences } \pi_2, \ldots, \pi_h \text{ remain and are orthogonal} \bigr)}\\
        \geq & \maxParaNOM(t,\sigma,t_{v'}) +  m_{\pi_2} + \ldots +  m_{\pi_h} \\
        \textcolor{blue}{\big\Downarrow} & \quad \textcolor{blue}{\bigl( \text{by the previous inequation} \bigr)}\\
        \geq & \maxParaNOM(t,\sigma,t_v) + m_{\pi_1} - 1 +  m_{\pi_2} + \ldots +  m_{\pi_h} \\
        \textcolor{blue}{\big\Downarrow} & \quad \textcolor{blue}{\bigl( m_{\pi_1} +  m_{\pi_2} + \ldots +  m_{\pi_h} = k \bigr)}\\
        = & \maxParaNOM(t,\sigma,t_v)  + k- 1
    \end{align*}
\end{myproof}